\documentclass[11pt,letterpaper]{article}

\usepackage{graphicx}

\begin{document}

\title{\large REDUCTION OF SPURIOUS VELOCITY\\
IN FINITE DIFFERENCE LATTICE BOLTZMANN MODELS\\
FOR LIQUID - VAPOR SYSTEMS}

\author{\normalsize Artur CRISTEA and
Victor SOFONEA\footnote{Corresponding author.}\medskip\\
{\normalsize\it Laboratory for Numerical Simulation and Parallel Computing in
Fluid Mechanics}\\
{\normalsize\it Center for Fundamental and Advanced Technical Research,
Romanian Academy}\\
{\normalsize\it Bd.~Mihai Viteazul 24, R -- 1900 Timi\c soara,
Romania}\medskip\\
{\normalsize E-mail: \{f1astra,sofonea\}@acad-tim.utt.ro}}

\maketitle

\abstract{The origin of the spurious interface velocity in finite
difference lattice Boltzmann models for liquid - vapor systems
is related to the first order upwind scheme used to compute
the space derivatives in the evolution equations. A correction
force term is introduced to eliminate the spurious velocity. The
correction term helps to recover sharp interfaces and sets the phase
diagram close to the one derived using the Maxwell construction.medskip\\


\section*{\normalsize
1. General description of Finite Difference Lattice Boltzmann models}

Lattice Boltzmann (LB) models \cite{a1,a2,a3,a4}
provide an alternative to current simulation
methods in computational fluid dynamics. These
models are based on the physics at
the mesoscopic scale, so that the macroscopic phenomena are recovered without
solving the equations of continuous media mechanics. The starting point of any
LB model is the Boltzmann equation \cite{a5}
\begin{equation}
\left(\,\frac{\partial}{\,\partial t\,}\,+\,{\bf{v}}\cdot\nabla\,+\,
\frac{\,{\mathbf{F}}\,}{\,m\,}\,\cdot\,\nabla_{{\mathbf{v}}}\,\right)\,
f\,=\,\left(\,\frac{\,\partial f\,}{\,\partial t\,}\,
\right)_{collisions}
\label{ecb}
\end{equation}
Here $f \equiv f({\mathbf{r}},{\mathbf{v}},t)$ is the distribution function
of fluid particles (supposed to be identical, with mass $m$),
${\mathbf{v}}$ is the particle
velocity and ${\mathbf{F}} = {\mathbf{F}}({\mathbf{r}},t)$ is the
local force acting on fluid particles.
The collision term in the Boltzmann equation (\ref{ecb}) is usually linearized
using the Bhatnagar-Gross-Krook (BGK) approximation \cite{a6}
after introducing a relaxation time $\tau$:
\begin{equation}
\left(\,\frac{\,\partial f\,}{\,\partial t\,}\,\right)_{collisions}\,=\,
-\,\frac{1}{\tau}\,\left[\,f({\bf{r}},{\bf{v}},t)\,-\,
f^{eq}({\bf{r}},{\bf{v}},t)\,
\right]
\label{liniar}
\end{equation}

The equilibrium distribution function
$f^{eq} \equiv f^{eq}({\mathbf{r}},{\mathbf{v}},t)$ which
appears in eq.~(\ref{liniar}) is the Maxwell - Boltzmann distribution function:
\begin{equation}
f^{eq}({\bf{r}},{\bf{v}},t)\,=\,n({\bf{r}},t)\,\left(\,\frac{m}{\,2\pi k_BT\,}
\,\right)^{1/2}\,\exp\left\{\,-\,\frac{m}{2k_BT\,}\cdot
\left[{\bf{v}}-{\bf{u}}({\bf{r}},t)\right]^2\,\right\}
\label{emaxboltz}
\end{equation}
where $k_B$ is the Boltzmann constant, $T$ is the absolute temperature of
the system,
\begin{equation}
n({\bf{r}},t)\,=\,\int\,f({\bf{r}},{\bf{v}},t)\,{\mathrm{d}}{\mathbf{v}}
\label{crho}
\end{equation}
is the particle number density and
\begin{equation}
{\bf{u}}({\bf{r}},t)\,=\,\frac{1}{\,n({\bf{r}},t)\,}\,
\int\,{\bf{v}}\cdot f({\bf{r}},{\bf{v}},t)\,{\mathrm{d}}{\mathbf{v}}
\label{cvel}
\end{equation}
is the local fluid velocity \cite{a7}.
We assume that the system is not too far from the equilibrium state and
get \cite{a8,a9}
\begin{equation}
\nabla_{\bf{v}}f({\bf{r}},{\bf{v}},t)\,\simeq\,
\nabla_{\bf{v}}f^{eq}({\bf{r}},{\bf{v}},t)\,=\,-\,\frac{m}{\,k_BT\,}\,
\left[\,{\bf{v}}\,-\,{\bf{u}}({\bf{r}},t)\,\right]\,
f^{eq}({\bf{r}},{\bf{v}},t)
\label{gradequiv}
\end{equation}

Recent investigations \cite{a10,a11,a12}
resulted in a general procedure to construct
lattice Boltzmann models for single - component fluids.
After discretization of the phase space \cite{a1,a2,a3,a4},
the distribution functions are defined only in the nodes ${\mathbf{x}}$
of a discrete
lattice ${\mathcal{L}}$ in the one - $(1D)$, two - $(2D)$ or three -
dimensional $(3D)$ space, while the velocities are reduced to a discrete set
$\{{\mathbf{e}}_i\},\,i=0,\,1,\ldots{\mathcal{N}}$. The elements of the
$1D$, $2D$ and $3D$ velocity sets $\{{\mathbf{e}}_i\}$ currently used in the LB
literature \cite{a3,a4,a10,a13} are expressed using the propagation speed
$c\,=\,\sqrt{k_BT/\chi m}$, where $\chi$ is a constant specific to each set.
Following the discretization
procedure, the Boltzmann equation (\ref{ecb}) is replaced by the set of
${\mathcal{N}}$ equations
\begin{eqnarray}
\partial_{t} f_i({\mathbf{x}},\,t)\,+\,
{\mathbf{e}}_i\,\cdot\,\nabla f_i({\mathbf{x}},\,t) & = &
- \, \frac{\,1\,}{\,\tau\,}\,\left[\, f_i({\mathbf{x}},t)
\,-\, f^{eq}_i({\mathbf{x}},t)\,\right]\nonumber \\
& + & \frac{\,1\,}{\,\chi c^2\,}\,f_i^{eq}({\mathbf{x}},t)\,
\left[\,{\mathbf{e}}_i\,-\,{\mathbf{u}}({\mathbf{x}},\,t)\,\right]\,
\cdot\,{\mathbf{F}}({\mathbf{x}},\,t) \rule{0mm}{7mm}\nonumber \\
& & (i\,=\,0,\,1,\,\ldots\,{\mathcal{N}})
\rule{0mm}{6mm}
\label{fdlbeqadimfin}
\end{eqnarray}
where the distribution functions $f_i({\mathbf{x}},\,t)$ express the
probability of finding at node ${\mathbf{x}}\in{\mathcal{L}}$ a particle
having the velocity ${\mathbf{e}}_i$. The particle number density $n$
and the local velocity ${\mathbf{u}}$ are now expressed as
\begin{eqnarray}
n & = & n({\mathbf{x}},\,t)\,=\,\sum_{i=0}^{i={\mathcal{N}}}\,
f_i({\mathbf{x}},\,t)\,=\,\sum_{i=0}^{i={\mathcal{N}}}\,
f_i^{eq}({\mathbf{x}},\,t)
\label{nsum} \\
{\mathbf{u}} & = & {\mathbf{u}}({\mathbf{x}},\,t)\,=\,
\frac{\,1\,}{\,n({\mathbf{x}},t)\,}\,
\sum_{i=0}^{i={\mathcal{N}}}\,
{\mathbf{e}}_i f_i({\mathbf{x}},\,t)\,=\,
\frac{\,1\,}{\,n({\mathbf{x}},t)\,}\,
\sum_{i=0}^{i={\mathcal{N}}}\,
{\mathbf{e}}_i f_i^{eq}({\mathbf{x}},\,t)
\rule{0mm}{7mm}\label{usum} 
\end{eqnarray}
while the equilibrium distribution functions
in (\ref{fdlbeqadimfin}) are given as series expansion in the local velocity:
\begin{equation}
f^{eq}_i \, = \, f^{,eq}_i({\mathbf{x}},\,t)\,=\,
w_i\,n\,\left[\,1\,+\,\frac{\,{\mathbf{e}}_i\,\cdot\,
{\mathbf{u}}\,}{\,\chi\,c^2\,}\,+\,
\frac{\,({\mathbf{e}}_i\,\cdot\,{\mathbf{u}})^2\,}
{\,2\,\chi^2\,c^4\,}\,-\,\frac{\,{\mathbf{u}}\,\cdot\,
{\mathbf{u}}\,}{\,2\,\chi\,c^2\,}\,\right]
\label{fdeqf}
\end{equation}

In the one - dimensional $D1Q3$ model \cite{a4}, $\chi\,=\,1/3$,
${\mathcal{N}}=2$ and the velocities ${\bf{e}}_i$ are given by:
\begin{eqnarray}
{\mathbf{e}}_i\,=\,\left\{\begin{array}{rl}
0\quad & (i\,=\,0)\\
c\quad & (i\,=\,1)\,\rule{0mm}{5mm}\\
-c\quad & (i\,=\,2)\,\rule{0mm}{5mm}\\
\end{array}
\right.
\label{speeds1d}
\end{eqnarray}
while the weight factors $w_i$ in (\ref{fdeqf}) are
\begin{equation}
w_i\,=\,\left\{\,\begin{array}{ll}
4/6 & (i\,=\,0)\\
1/6 & (i\,=\,1,\,2) \rule{0mm}{5mm}
\end{array}\right.
\end{equation}
When using a square lattice in the $2D$ space (D2Q9 model \cite{a14}),
$\chi\,=\,1/3$,
${\mathcal{N}}\,=\,8$ and the velocities ${\mathbf{e}}_i$ are given by
\begin{equation}
{\mathbf{e}}_i\,=\,\left\{\,
\begin{array}{ll}
0 & (i\,=\,0)\\
\left[\,\cos\frac{\,(i-1)\pi\,}{2}\, ,\,\sin\frac{\,(i-1)\pi\,}{2}\,
\right]\,c\qquad & (i\,=\,1,\dots 4) \rule{0mm}{8mm}\\
\left[\,\cos\left(\frac{\,\pi\,}{\,4\,}\,+\,\frac{\,(i-5)\pi\,}{2}\right)\,
,\,\sin\left(\,\frac{\,\pi\,}{\,4\,}\,+\,\frac{\,(i-5)\pi\,}{2}\,\right)\,
\right]\,\sqrt{2}\,c & (i\,=\,5,\dots 8)\rule{0mm}{8mm}
\end{array}
\right.
\label{ec9}
\end{equation}
while the weight factors are
\begin{equation}
w_i\,=\,\left\{\,\begin{array}{ll}
4/9 & (i\,=\,0)\\
1/9 & (i\,=\,1,\dots 4) \rule{0mm}{5mm}\\
1/36 & (i\,=\,5,\dots 8)\rule{0mm}{5mm}
\end{array}\right.
\end{equation}

The Cartesian projections $e_{i\alpha}\,\equiv\,({\mathbf{e}}_i)_\alpha$
$(i\,=\,1,\,2,\ldots ,{\mathcal{N}};\quad \alpha\,=\,x,\,y,\ldots)$ of
the velocity vectors ${\mathbf{e}}_i$ satisfy the relations
\begin{eqnarray}
\sum_i\,w_i e_{i\alpha} & = & 0 \nonumber\\
\sum_i\,w_i e_{i\alpha} e_{i\beta} & = & \chi c^2\delta_{\alpha\beta}
\rule{0mm}{5mm}\nonumber\\
\sum_i\,w_i e_{i\alpha} e_{i\beta}  e_{i\gamma} & = &
 0 \rule{0mm}{5mm}\nonumber\\
\sum_i\,w_i e_{i\alpha} e_{i\beta} e_{i\gamma} e_{i\delta} & = &
\chi^2 c^4 (\delta_{\alpha\beta}\delta_{\gamma\delta}\,+\,
\delta_{\beta\gamma}\delta_{\delta\alpha}\,+\,
\delta_{\alpha\gamma}\delta_{\beta\delta})
\rule{0mm}{5mm} \label{sume}
\end{eqnarray}
The following sums are easily computed
using the definition (\ref{fdeqf}) of the equilibrium distribution functions:
\begin{eqnarray}
\sum_i f_{i}^{eq} & = & n \nonumber\\
\sum_i e_{i\alpha} f_{i}^{eq}  & = & n u_\alpha \nonumber\\
\sum_i e_{i\alpha} e_{i\beta} f_{i}^{eq} & = & n\,[\,\chi c^2
\delta_{\alpha\beta} \,+\, u_\alpha u_\beta\, ]\nonumber\\
\sum_i e_{i\alpha} e_{i\beta} e_{i\gamma} f_{i}^{eq} & = &
n \chi c^2\,[\,\delta_{\alpha\beta}u_\gamma\,+\,
\delta_{\beta\gamma}u_\alpha\,+\,\delta_{\gamma\alpha}u_\beta\,]
\label{esumfeq}
\end{eqnarray}
Here $u_\alpha$ ($\alpha\,=\,x,\,y$) are the Cartesian components of the
local velocity ${\mathbf{u}}$.

The set of phase space discretized LB equations (\ref{fdlbeqadimfin})
for the distribution functions  $f_i\,=\,f_i({\mathbf{x}},\,t)$ may be solved
numerically using an appropriate finite difference scheme defined
on the lattice ${\mathcal{L}}$. When using a scheme based on the
characteristics line, the
forward Euler difference is used to compute the time derivative while
the first - order upwind scheme may be used for the space derivative
\cite{a13}
as usually done in classical LB models \cite{a3,a4}. These schemes give
the following updating procedure
for the distribution function \cite{a13,a15,a16}
\begin{eqnarray}
f_i({\mathbf{x}},t+\delta t) & = & f_i({\mathbf{x}},t)\,+\,
\frac{\,c\delta t\,}{\,\delta s\,}\,\left[\,f_i({\mathbf{x}},t)\,-\,
f_i({\mathbf{x}}-\,\delta s{\mathbf{e}}_i/c,t)\,\right]
\label{upwindscheme} \\
& - & \frac{\,\delta t\,}{\,\tau\,}\,\left[\,f_i({\mathbf{x}},t)
\,-\, f^{eq}_i({\mathbf{x}},t)\,\right]\,+\,
\frac{\,\delta t\,}{\,\chi c^2\,}\,f_i^{eq}({\mathbf{x}},t)\,
\left[\,{\mathbf{e}}_i\,-\,{\mathbf{u}}({\mathbf{x}},\,t)\,\right]\,
\cdot\,{\mathbf{F}}({\mathbf{x}},\,t) \rule{0mm}{5mm} \nonumber
\end{eqnarray}
where $\delta t$ is the time step and $\delta s$ is the lattice spacing.

\section*{\normalsize 2. Dimensionless equations and the force term}

Let $l_R$, $t_R$, $n_R$, $T_R$, $c_R$ and $a_R$ the reference quantities for
length, time, particle number density, temperature, speed and acceleration.
In order to preserve the form (\ref{fdlbeqadimfin}) of the LB evolution
equations, the reference quantities should satisfy the following relations
\begin{eqnarray}
\frac{\,t_R c_R\,}{\,l_R\,}\,=\,1 \nonumber\\
\frac{\,a_R t_R\,}{\,c_R\,}\,=\,1 \rule{0mm}{6mm}
\label{refcond}
\end{eqnarray}
Since we will refer further to a van der Waals fluid, a natural choice
for $n_R$, $T_R$ and $c_R$ may be the values corresponding to a mole of fluid
at the critical point \cite{a17},
$n_R\,=\,N_A/V_{mc}$, $T_R\,=\,T_c$ and $c_R\,=\,\sqrt{k_B T_c/m}$,
respectively. 
Here $N_A$ is Avogadro's number, $V_{mc}$ is the molar volume at the
critical point and $T_c$ is the critical temperature.
With this choice of the reference quantities, the dimensionless
propagation speed is
\begin{equation}
c\,=\,\sqrt{T/\chi}
\label{adimc}
\end{equation}
where $T$ is the dimensionless
temperature (note that $T\,=\,1$ at the critical point).
The other reference quantities may be derived from Eqs.~(\ref{refcond}).
if we choose the characteristic system size as reference length.

The fact that the dimensionless
propagation velocity $c$ (\ref{adimc}) should not necessarily equal $1$
is a characteristics of the Finite Difference LB models \cite{a18}, where the
propagation velocity is no longer related to the lattice spacing
as in standard (classical) LB models \cite{a3,a4}. 
Comparison of fluid velocity profiles at different temperatures becomes
easier with the definition (\ref{adimc}) associated to the expression
(\ref{usum}) of the local fluid velocity.

The local force acting on a fluid particle is the sum of two terms:
\begin{equation}
{\mathbf{F}}({\mathbf{x}},\,t)\,=\,{\mathbf{F}}^{\varphi}\,+\,
{\mathbf{F}}^{\sigma}
\label{forcesum}
\end{equation}
The first term
${\mathbf{F}}^{\varphi}$ accounts for phase separation
in the van der Waals fluid,
while the second one ${\mathbf{F}}^{\sigma}$ generates the
surface tension at the interface between phases. The
expression of ${\mathbf{F}}^{\varphi}$ and ${\mathbf{F}}^{\sigma}$
is adopted from the literature \cite{a9}
\begin{eqnarray}
{\mathbf{F}}^{\varphi} & = & \frac{\,1\,}{\,\rho\,}\,\nabla\,
(\,-\,p_w\,+\,\chi c^2\rho\,) \\
{\mathbf{F}}^{\sigma} & = & \,\kappa\,\nabla(\nabla^2\rho)
\rule{0mm}{6mm}
\end{eqnarray}
Here $p_w$ is the van der Waals pressure, $\chi c^2\rho$
is the dimensionless pressure of the ideal gas and
$\kappa$ is a parameter that controls the surface tension.
When using the reference pressure $p_R\,=\,m n_R c_R^2$, which is in
accordance to Eqs.~(\ref{refcond}), the dimensionless
van der Waals equation of state reads
\begin{equation}
p_w\,=\,\frac{\,\rho T\,}{\,3\,-\,\rho\,}\,-\,\frac{\,3\,}{\,8\,}\,\rho^2
\label{pvdw}
\end{equation}
Note that the dimensionless form (\ref{pvdw}) of the van der Waals
equation of state is different from the form used by other
authors \cite{a19,a20}.

\section*{\normalsize 3. Stationary case: origin of the
spurious velocity in the interface region}

Simulation results reported in this paper refer to the equilibrium
state of flat interfaces in liquid - vapor systems like the $2D$ one shown in
Figure 1. Density, velocity and pressure profiles, as well as the
phase diagram recovered in the stationary
(equilibrium) state are found to be identical when using both $D1Q3$
and $D2Q9$ models. For this reason, we refer to the $D1Q3$ model
on a $1D$ lattice with $N\,=\,100$ nodes (i.e., $\delta s\,=\,0.01$) when
reporting the simulation results, as follows.

\begin{figure}
\centerline{\includegraphics[width=100mm]{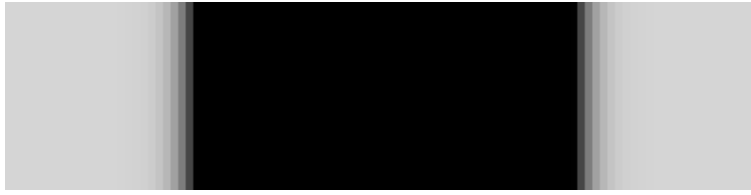}}
\caption{\normalsize
Plane interfaces in a $2D$ liquid - vapor system (the high density
phase is black). Lattice size: $100\times 25$ nodes; lattice spacing:
$\delta s\,=\,0.01$; time spacing: $\delta t\,=\,0.001$;
temperature: $T\,=\,0.70$; surface tension parameter: $k\,=\,0.00005$.}
\label{2dinterfaces}
\end{figure}

\begin{figure}
\centerline{\includegraphics[width=100mm]{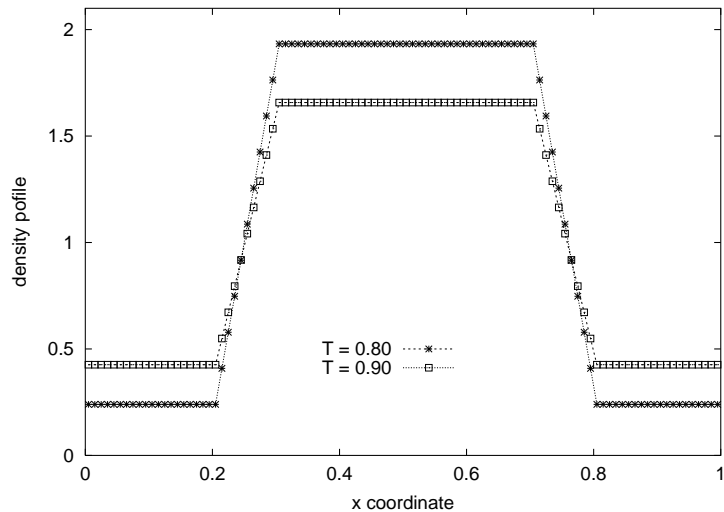}}
\caption{Initial density profiles for two values of the
dimensionless temperature.}
\label{rhoinitial}
\end{figure}

\begin{figure}
\centerline{\includegraphics[width=100mm]{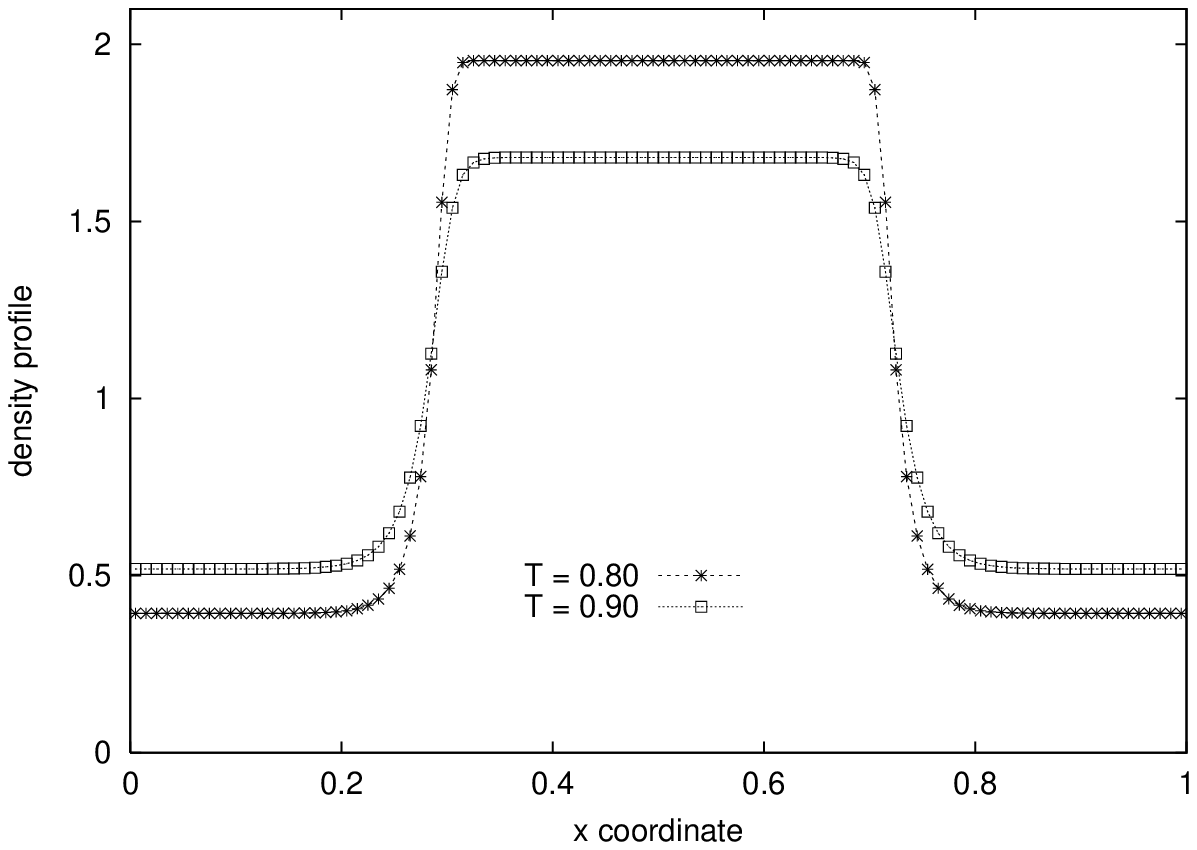}}

\centerline{a: $\kappa\,=\,0$}
\label{rhofinalupwind}
\caption{Equilibrium density profiles recovered with the first - order
upwind scheme (\protect{\ref{upwindscheme}})
for two values of the dimensionless temperature and different values of the
surface tension parameter $\kappa$.}
\end{figure}

\addtocounter{figure}{-1}
\begin{figure}
\centerline{\includegraphics[width=100mm]{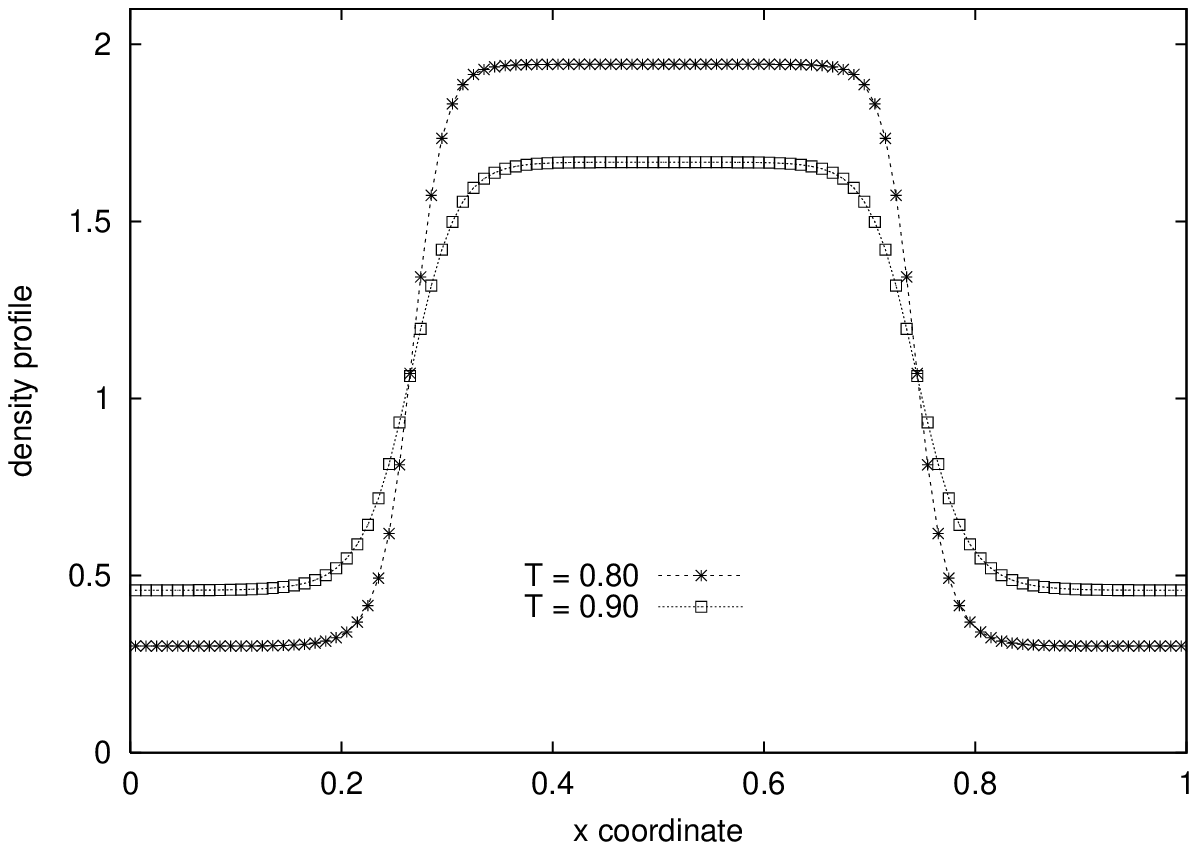}}

\centerline{b: $\kappa\,=\,0.00002$}

\centerline{\includegraphics[width=100mm]{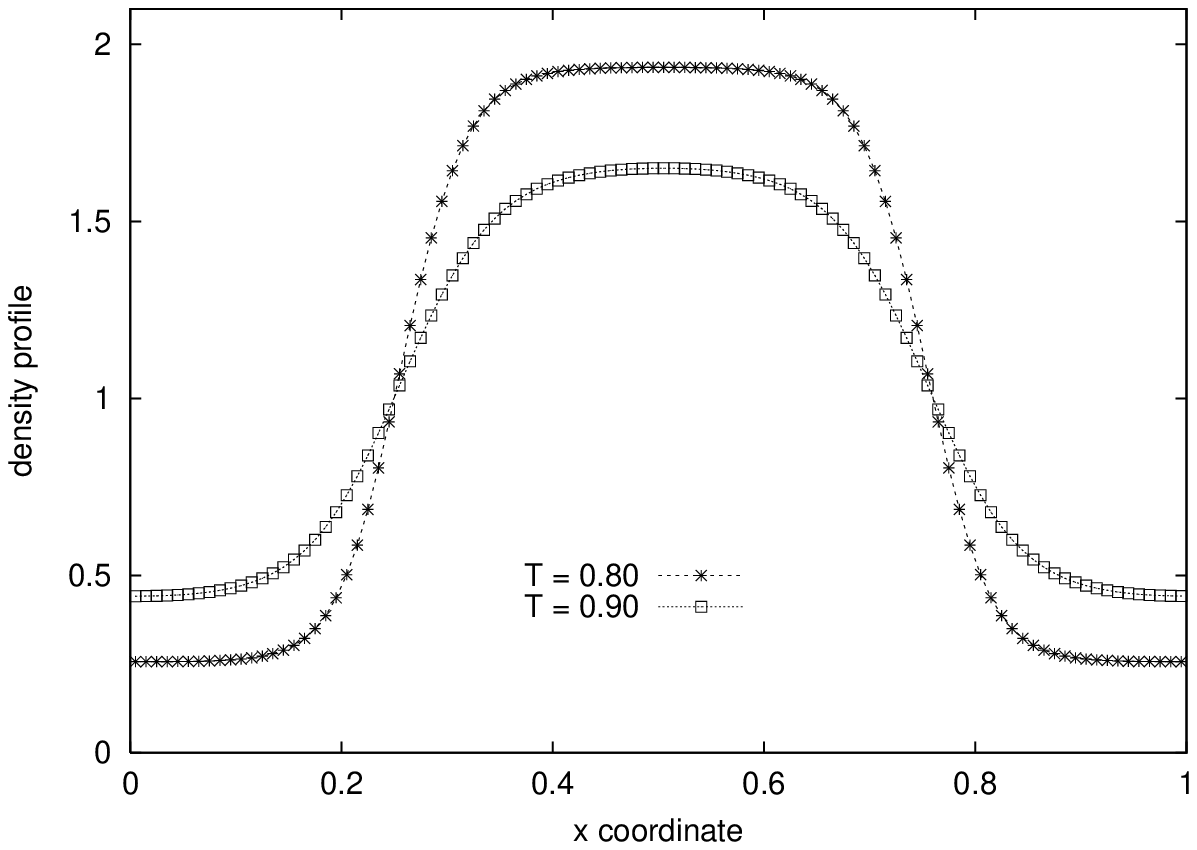}}

\centerline{c: $\kappa\,=\,0.0001$}
\label{rhofinalupwindcont}
\caption{{\emph{(cont'd)}}
Equilibrium density profiles recovered with the first - order
upwind scheme (\protect{\ref{upwindscheme}})
for two values of the dimensionless temperature and different values of the
surface tension parameter $\kappa$.}
\end{figure}

For each value of the temperature, the initial density profile
was set up to allow a transition region (Figure 2),
while the values of the initial
density of the liquid and vapor phases were computed using the Maxwell
construction \cite{a17}. This procedure (which avoids large density gradients)
helps the system to exhibit a stable behavior until the equilibrium state
is reached.
The dimensionless relaxation time was set to
$\tau\,=\,0.01$ and we always
performed 200,000 time steps ($\delta t\,=\,0.001$)
using the upwind finite difference scheme (\ref{upwindscheme})
to ensure the equilibrium state of the fluid system.

Simulations were done with three values of the parameter $\kappa$ which
controls the surface tension.
Figure 3 shows the resulting density profiles,
for $T\,=\,0.90$ and $T\,=\,0.80$. As expected \cite{a19,a20},
the interface width becomes larger as
$\kappa$ increases. But there is still a transition region between
the two phases for $\kappa\,=\,0$, when one would expect a sharp interface.
Velocity profiles derived in the equilibrium state
(Figure 4) show the existence of the spurious velocity (i.e.,
unphysical nonvanishing values of this quantity) in the interface region.
The magnitude of the spurious velocity becomes larger when decreasing the
temperature i.e., when the difference between the high and low density
phases increases. However, the spurious velocity in the interface region
reduces for larger values of the surface tension parameter $\kappa$, when
density gradients in the interface region become smaller.
Also, the interface
profiles of the van der Waals pressure (\ref{pvdw}) shown in Figure 5
become wider when the value of $\kappa$ is increased. For temperatures near
the critical point and large values of $\kappa$,
the left and right interfaces overlap one another. This is especially seen in
Figure 5c, where the pressure plateaux of both the high and low density
phase vanish for $T\,=\,0.90$.

The phase diagram of the liquid vapor system (Figure 6)
is strongly affected by the value of the surface tension parameter $\kappa$.
The liquid - vapor system becomes more stable (i.e., lower temperature
states may be reached) when using larger values of $\kappa$
which ensures smaller density gradients in the interface
region. Moreover,
when $\kappa$ is increased,
the computed values of the densities (especially in the vapor phase)
become closer to the theoretical values derived using the Maxwell construction.
The phase diagram becomes flawed near the critical point for
larger $\kappa$, but this is an effect of the overlapping of the 
right and left interfaces.

\begin{figure}
\centerline{\includegraphics[width=100mm]{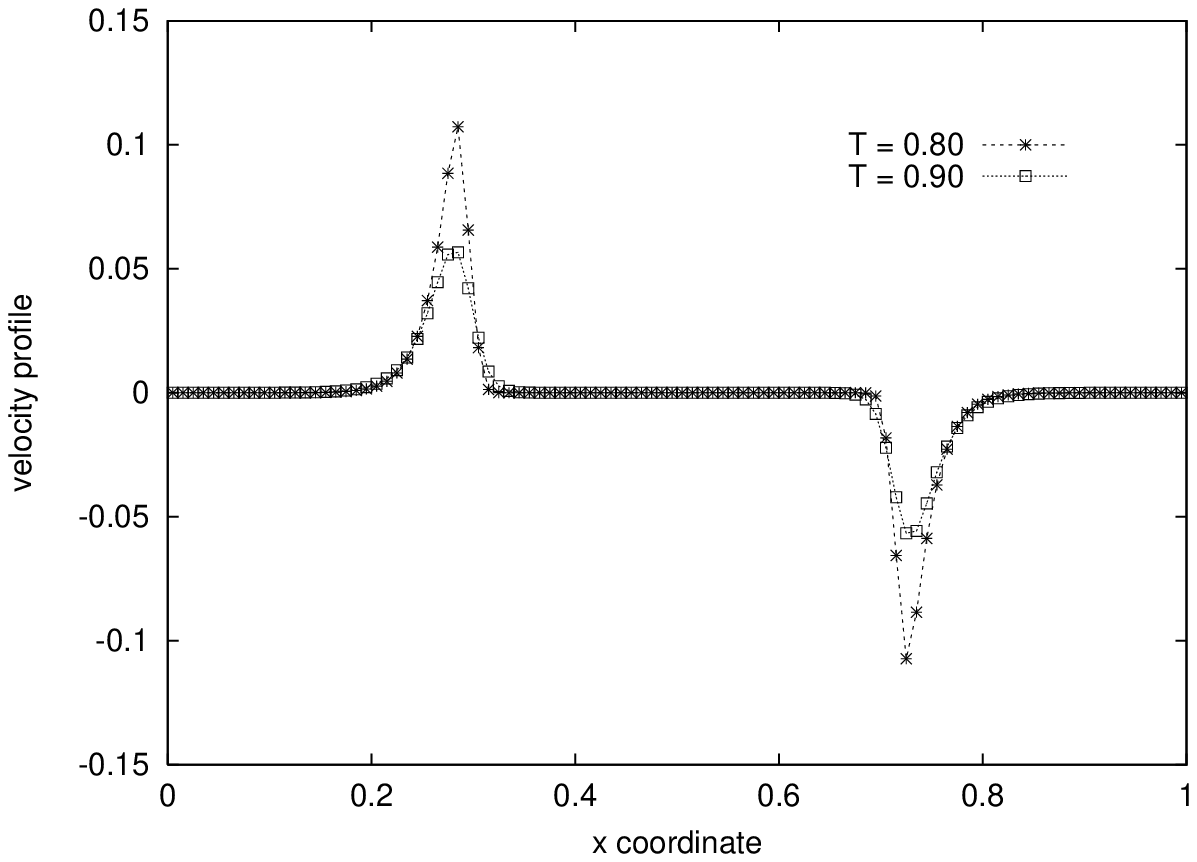}}

\centerline{a: $\kappa\,=\,0$}

\centerline{\includegraphics[width=100mm]{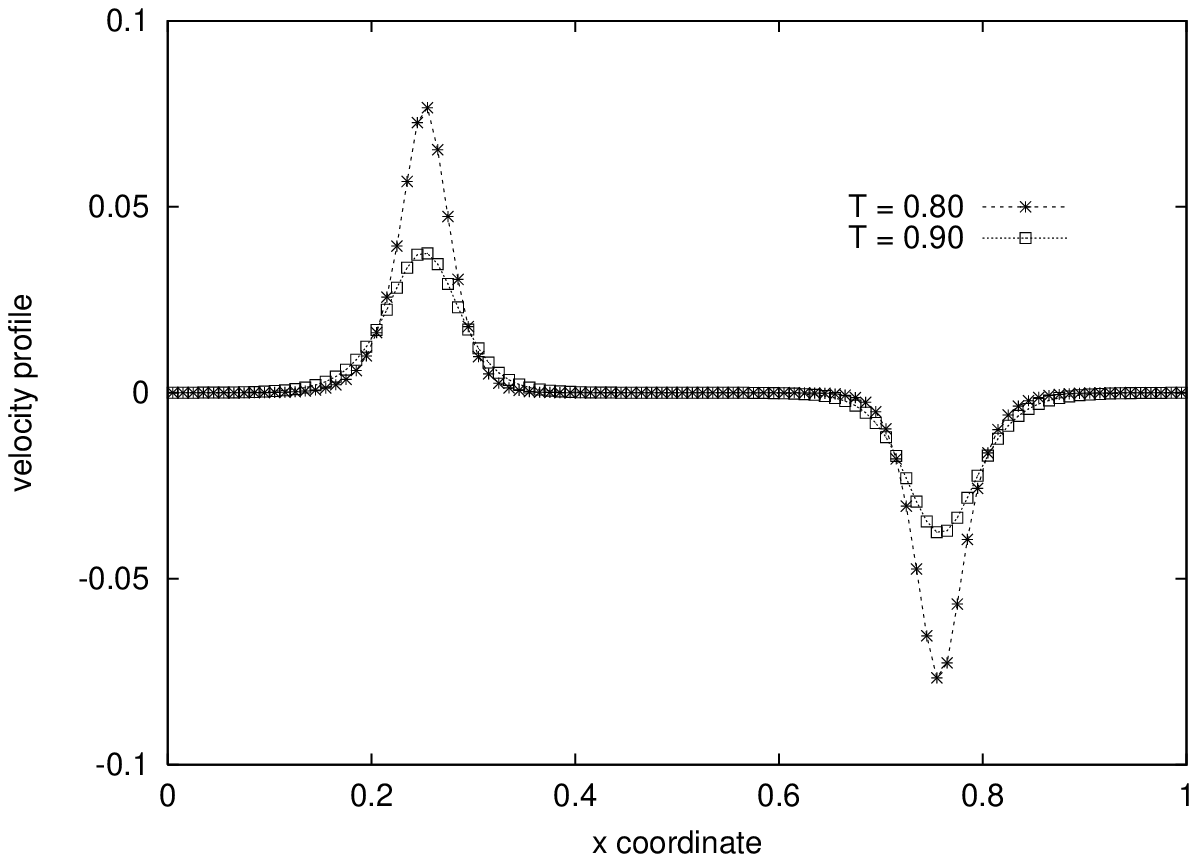}}

\centerline{b: $\kappa\,=\,0.00002$}

\label{velfinalupwind}
\caption{Equilibrium velocity profiles recovered with the first - order
upwind scheme (\protect{\ref{upwindscheme}})
for two values of the dimensionless temperature and different values of the
surface tension parameter $\kappa$.}
\end{figure}

\addtocounter{figure}{-1}
\begin{figure}
\centerline{\includegraphics[width=100mm]{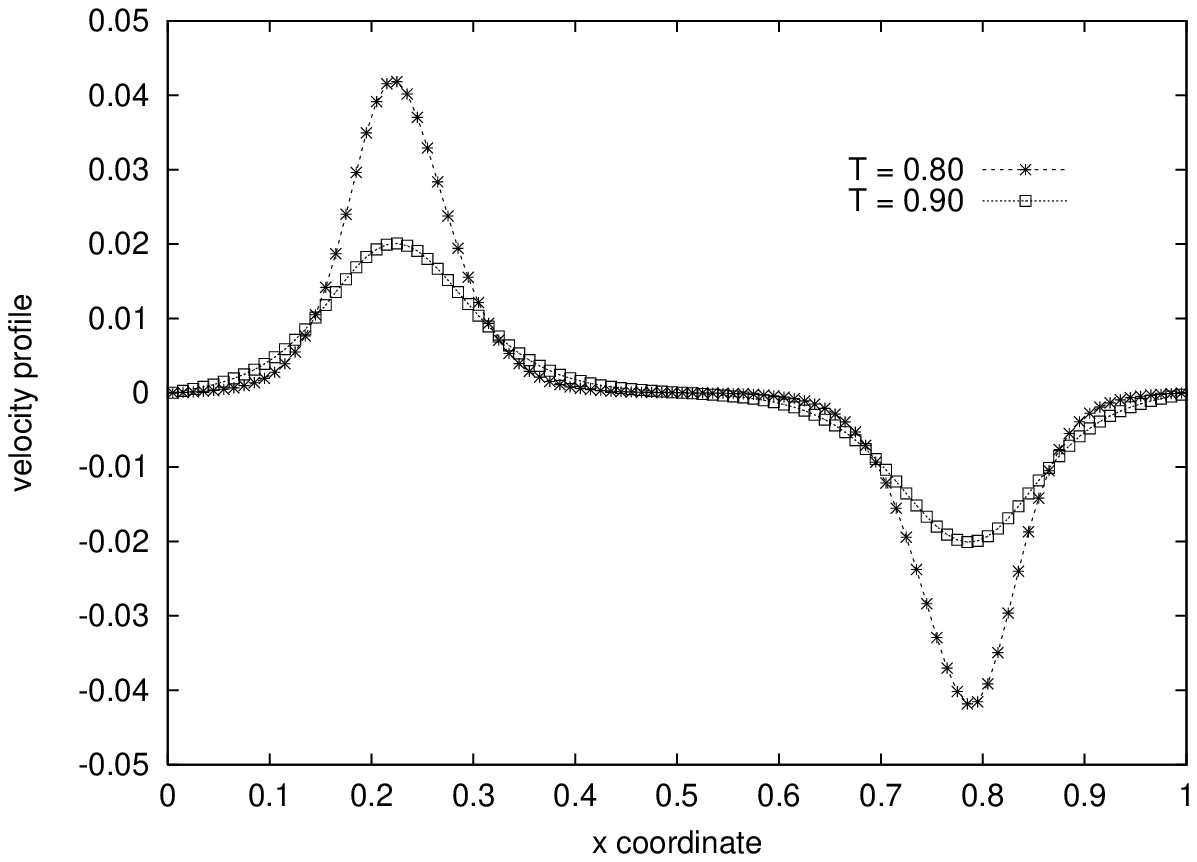}}

\centerline{c: $\kappa\,=\,0.0001$}

\label{velfinalupwindcont}
\caption{{\emph{(cont'd)}}
Equilibrium velocity profiles recovered with the first - order
upwind scheme (\protect{\ref{upwindscheme}})
for two values of the dimensionless temperature and different values of the
surface tension parameter $\kappa$.}
\end{figure}

\begin{figure}
\centerline{\includegraphics[width=100mm]{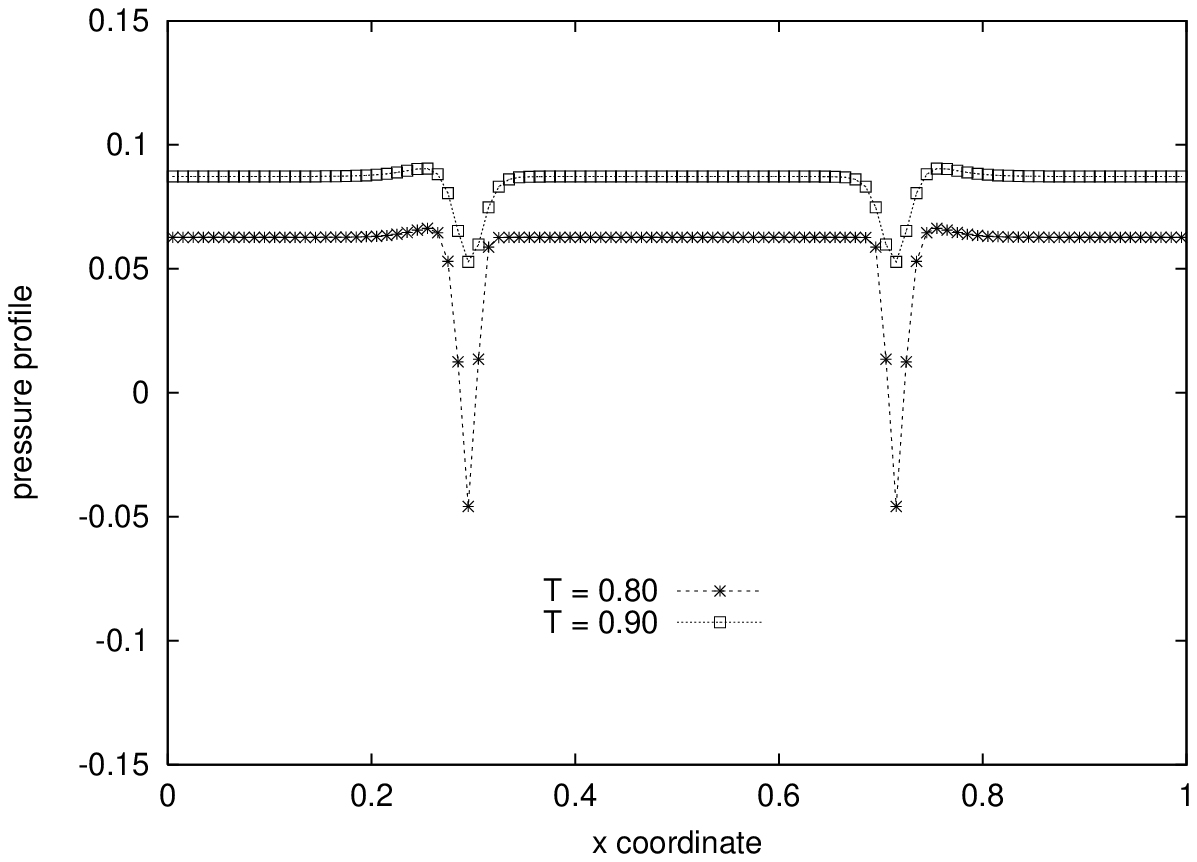}}

\centerline{a: $\kappa\,=\,0$}

\label{presfinalupwind}
\caption{Equilibrium pressure profiles recovered with the first - order
upwind scheme (\protect{\ref{upwindscheme}})
for two values of the dimensionless temperature and different values of the
surface tension parameter $\kappa$.}
\end{figure}

\addtocounter{figure}{-1}
\begin{figure}
\centerline{\includegraphics[width=100mm]{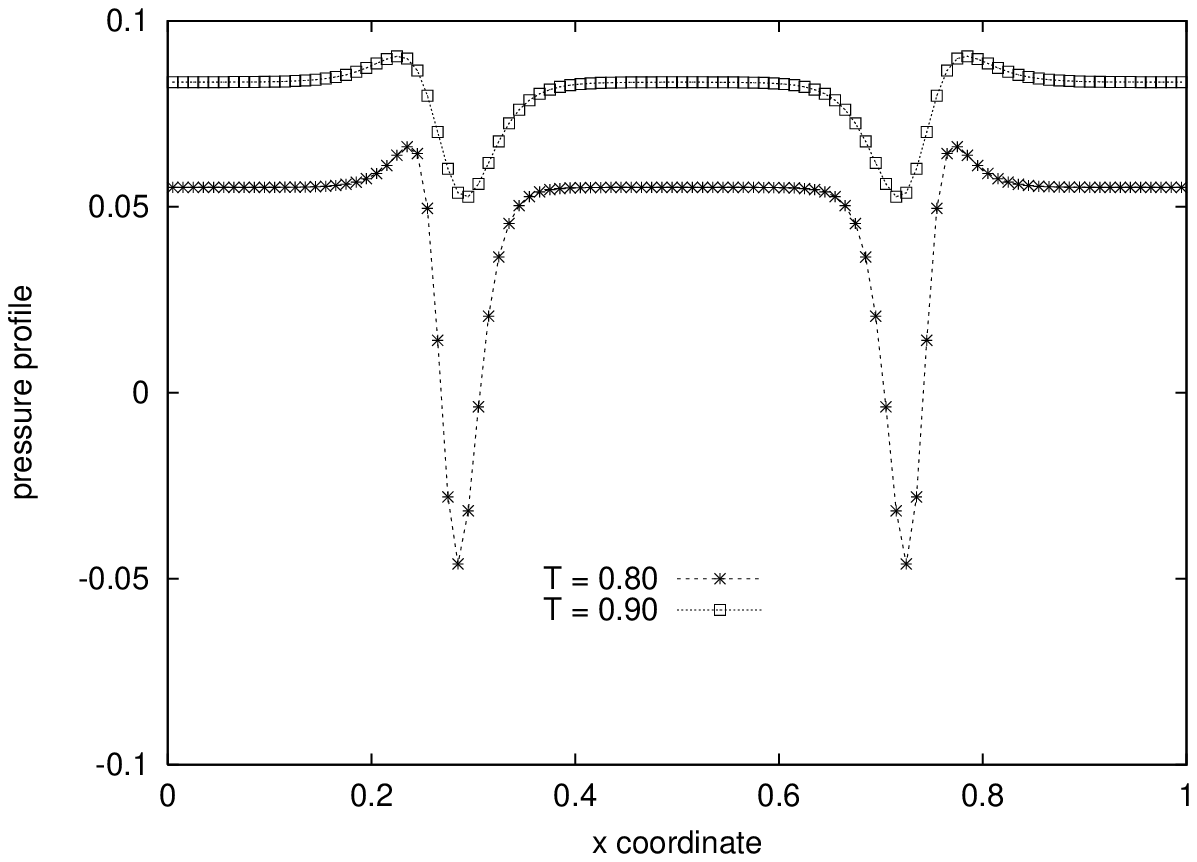}}

\centerline{b: $\kappa\,=\,0.00002$}

\centerline{\includegraphics[width=100mm]{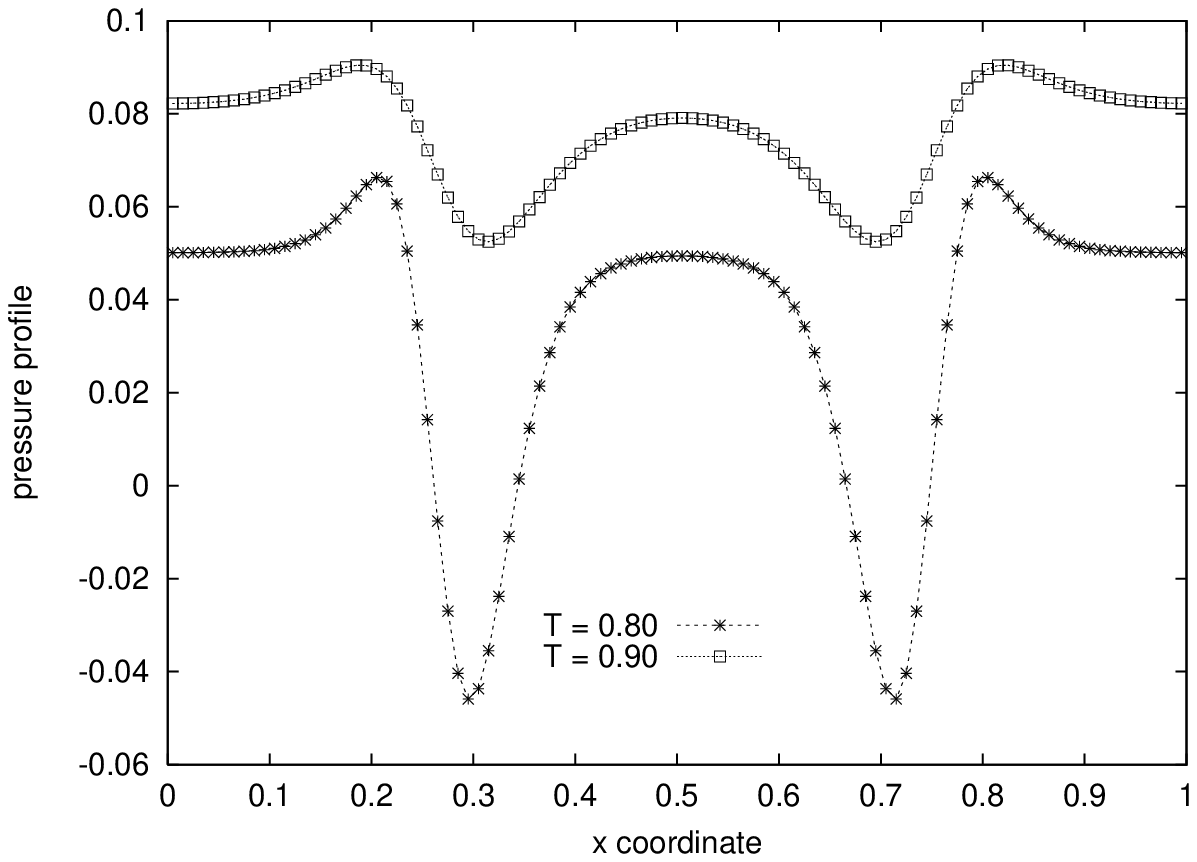}}

\centerline{c: $\kappa\,=\,0.0001$}

\label{presfinalupwindcont}
\caption{{\emph{(cont'd)}}
Equilibrium pressure profiles recovered with the first - order
upwind scheme (\protect{\ref{upwindscheme}})
for two values of the dimensionless temperature and different values of the
surface tension parameter $\kappa$.}
\end{figure}

\begin{figure}
\centerline{\includegraphics[width=100mm]{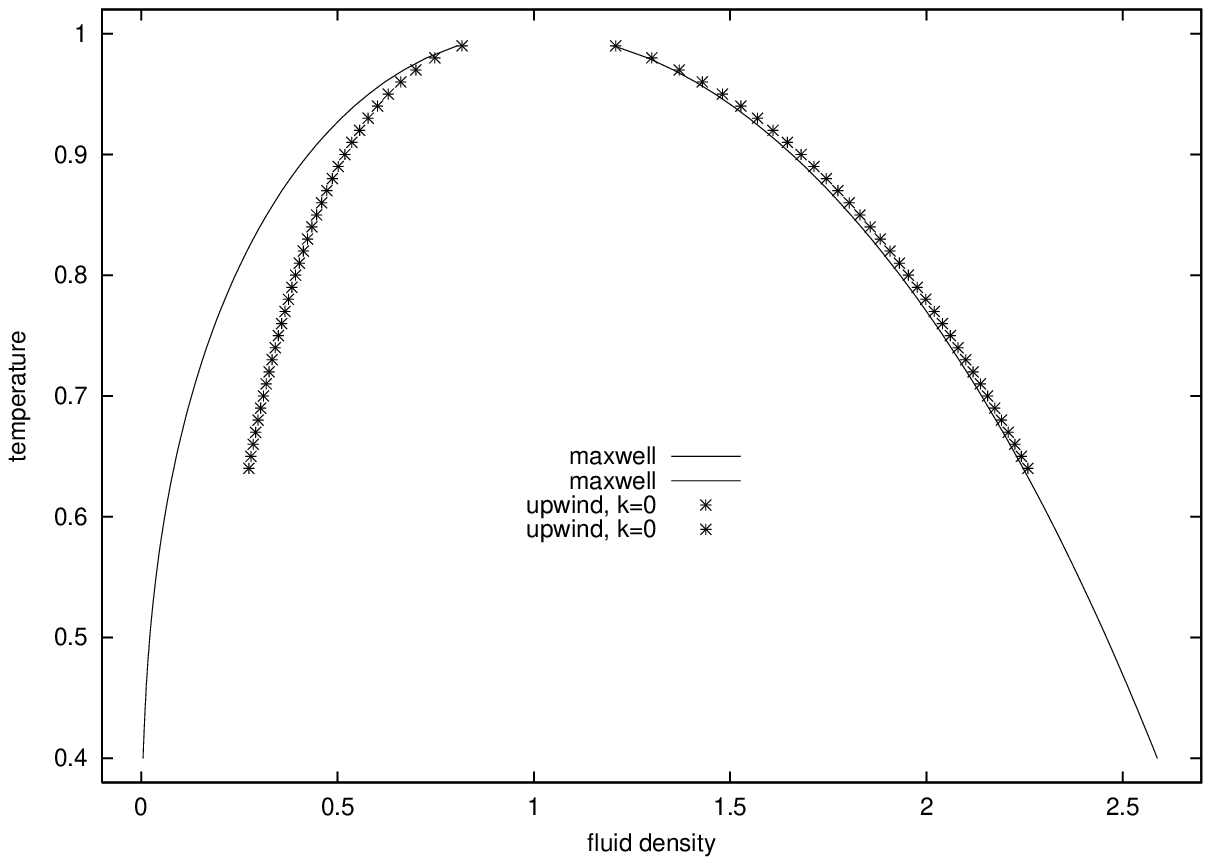}}

\centerline{a: $\kappa\,=\,0$}

\centerline{\includegraphics[width=100mm]{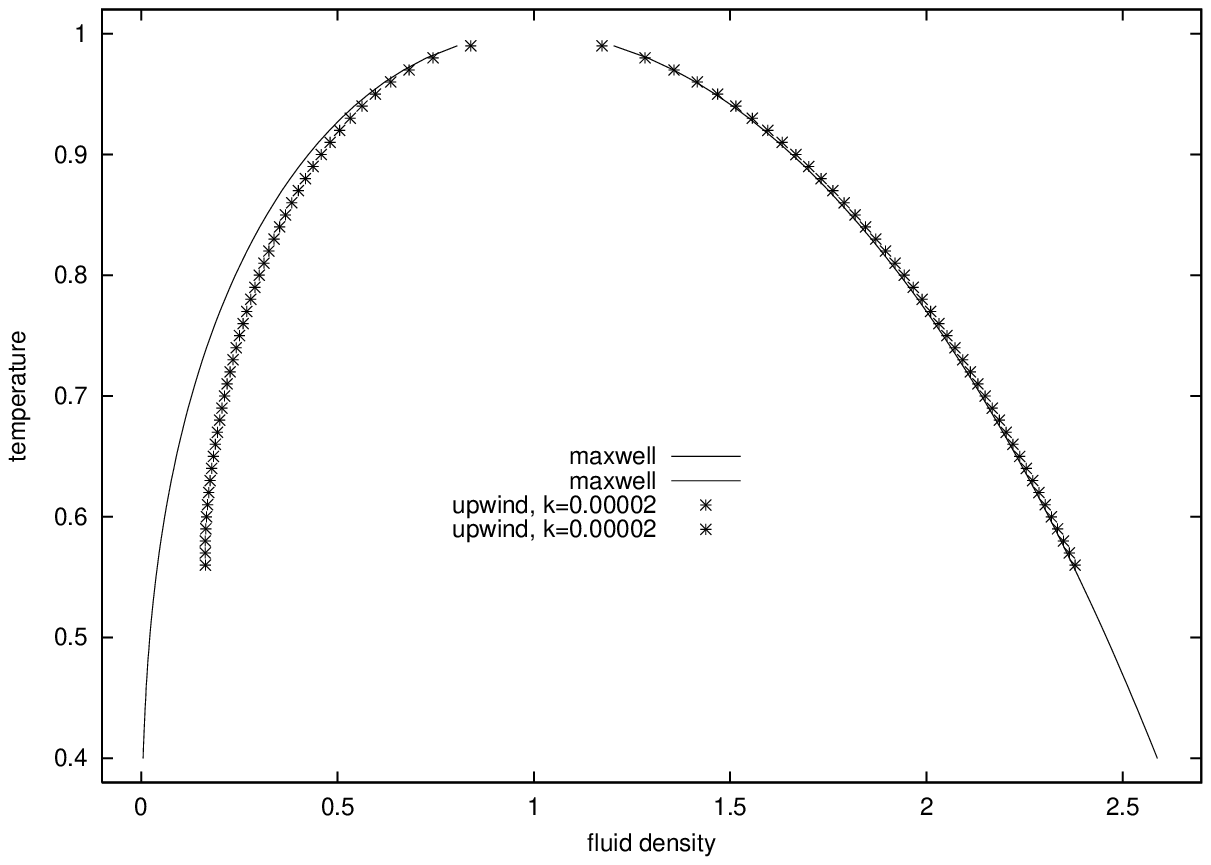}}

\centerline{b: $\kappa\,=\,0.00002$}

\label{phasediagupwind}
\caption{Phase diagrams recovered with the first - order
upwind scheme (\protect{\ref{upwindscheme}}) for different values of the
surface tension parameter $\kappa$.}
\end{figure}

\addtocounter{figure}{-1}
\begin{figure}
\centerline{\includegraphics[width=100mm]{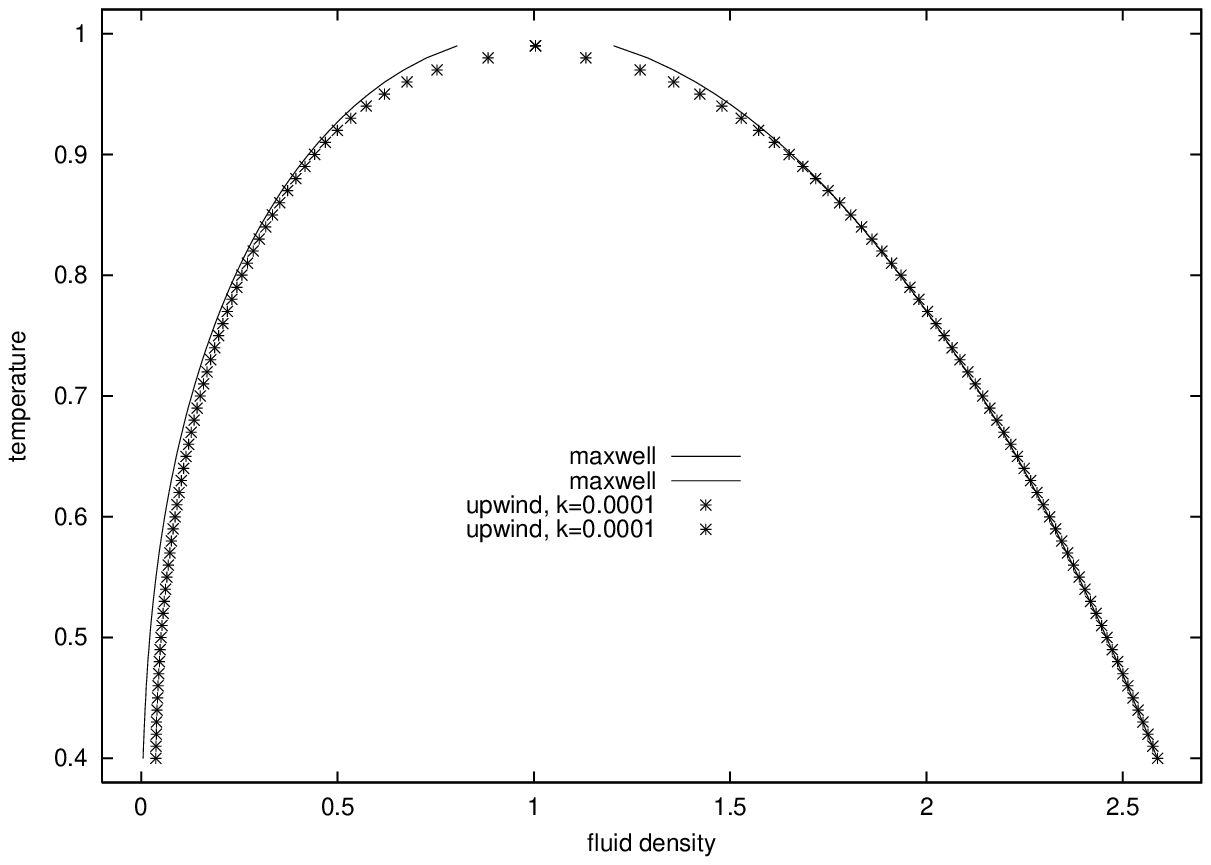}}

\centerline{c: $\kappa\,=\,0.0001$}

\label{phasediagupwindcont}
\caption{{\emph{(cont'd)}} Phase diagrams recovered with the first - order
upwind scheme (\protect{\ref{upwindscheme}}) for different values of the
surface tension parameter $\kappa$.}
\end{figure}

\begin{figure}
\centerline{\includegraphics[width=100mm]{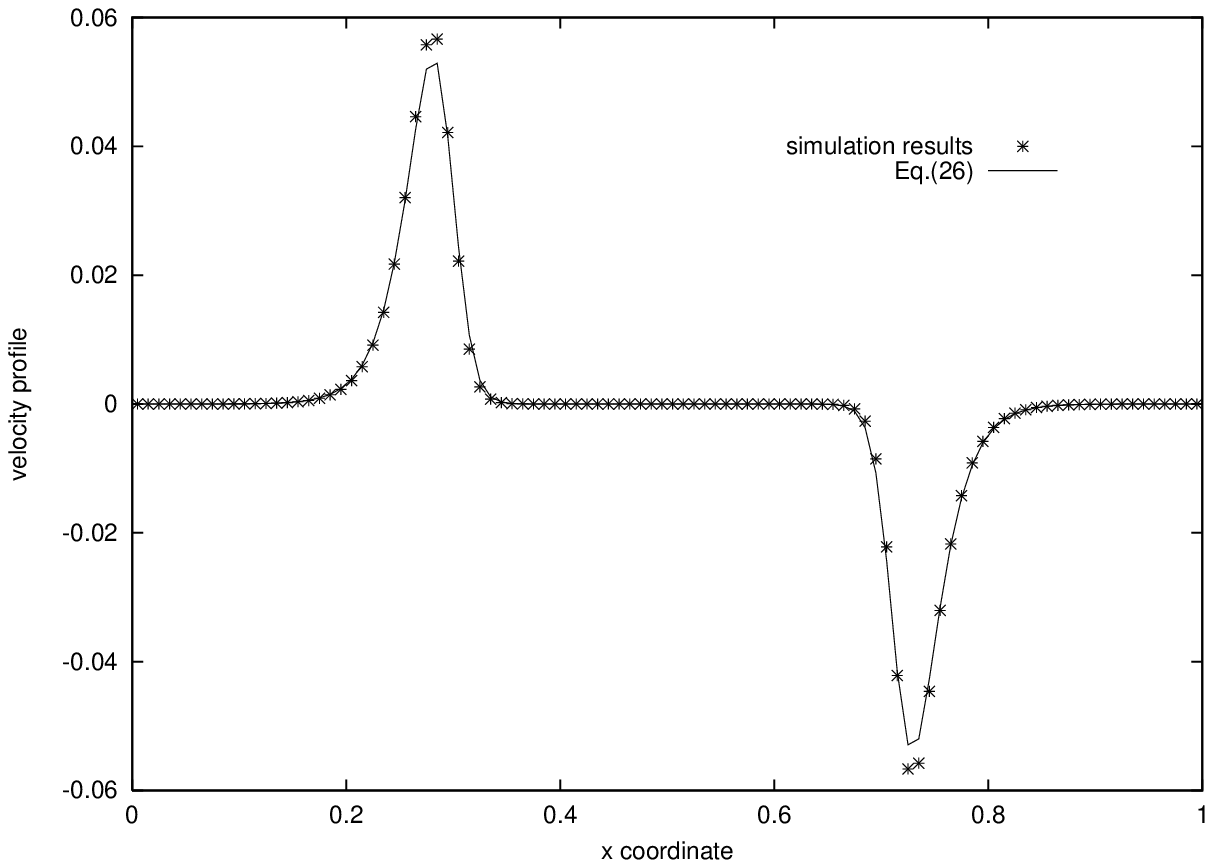}}

\centerline{a: $\kappa\,=\,0.00000$}

\label{figuspurious}
\caption{Velocity profile at $T\,=\,0.90$:
comparison between numerical results and Eq.~({\protect{\ref{uspurious}}})
for different values of the surface tension parameter $\kappa$.}
\end{figure}

\addtocounter{figure}{-1}
\begin{figure}
\centerline{\includegraphics[width=100mm]{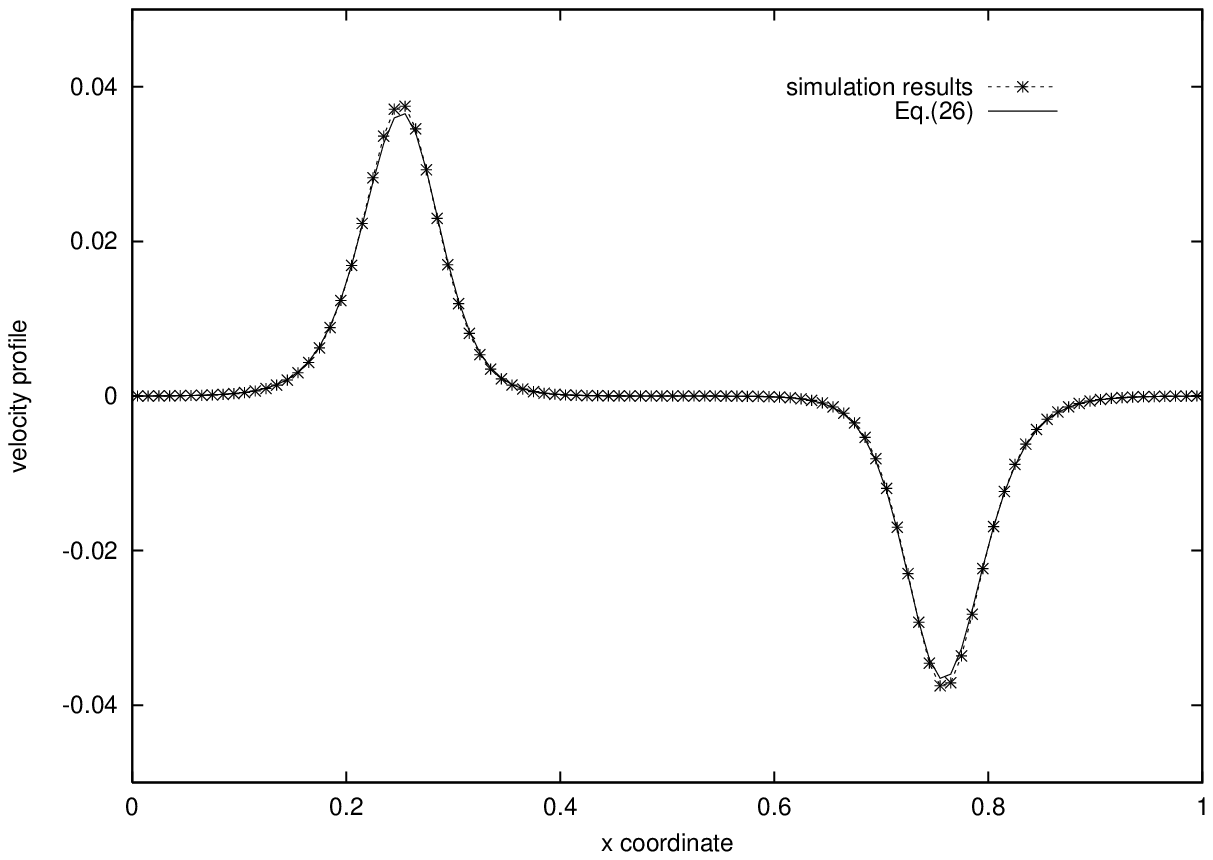}}

\centerline{b: $\kappa\,=\,0.00002$}

\centerline{\includegraphics[width=100mm]{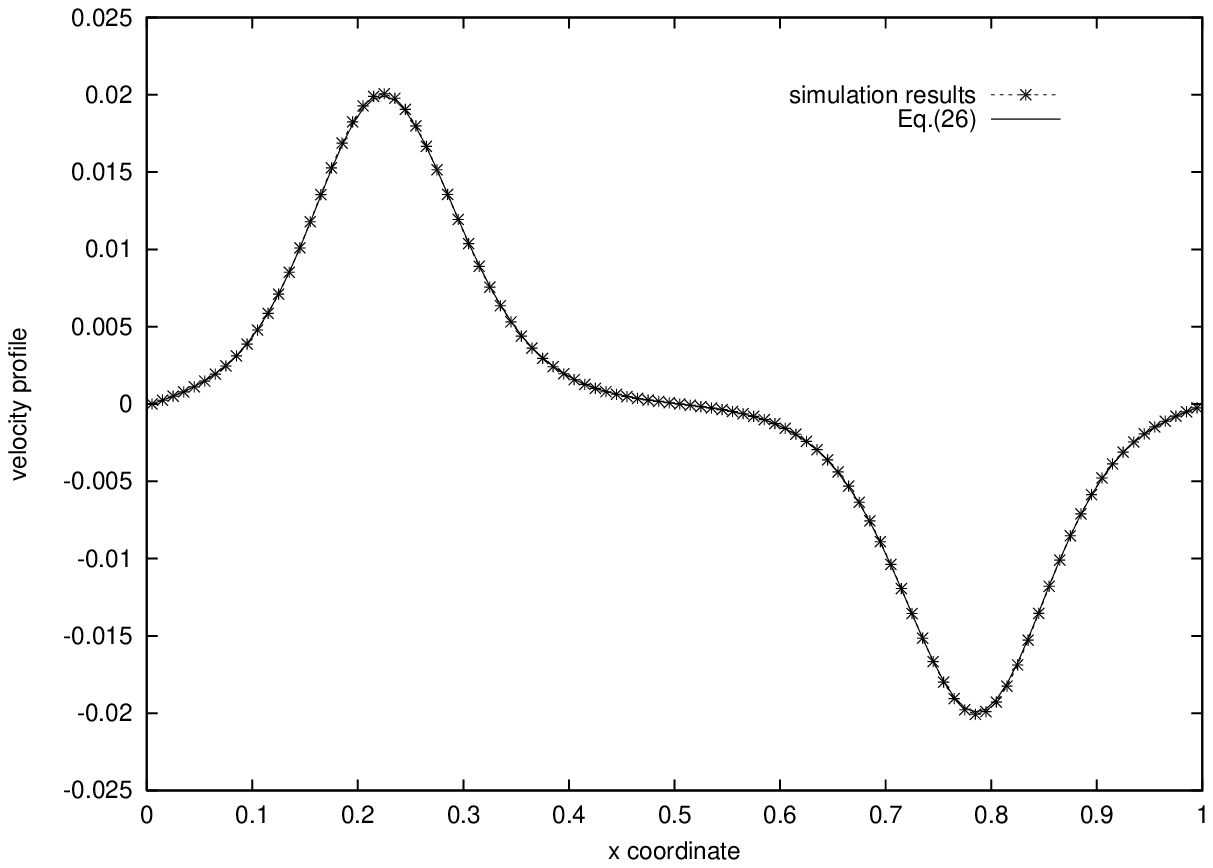}}

\centerline{c: $\kappa\,=\,0.0001$}

\caption{{\emph{(cont'd)}}
Velocity profiles at $T\,=\,0.90$:
comparison between numerical results and Eq.~({\protect{\ref{uspurious}}})
for different values of the surface tension parameter $\kappa$.}
\end{figure}

To account for the spurious interface velocity, we recall that
the {\emph{real}} LB equations \cite{a13} solved
using the upwind finite difference scheme in the stationary case are,
up to second order in the lattice spacing $\delta s$
\begin{eqnarray}
{\mathbf{e}}_i\,\cdot\,\nabla f_i({\mathbf{x}},\,t)\,-\,
\psi e_{i\beta}e_{i\gamma}\partial_\beta\partial_\gamma f_i({\mathbf{x}},t)
& = & - \, \frac{\,1\,}{\,\tau\,}\,\left[\, f_i({\mathbf{x}},t)
\,-\, f^{eq}_i({\mathbf{x}},t)\,\right]\nonumber \\
& + & \frac{\,1\,}{\,\chi c^2\,}\,f_i^{eq}\,
\left[\,e_{i\beta}\,-\,u_\beta({\mathbf{x}},\,t)\,\right]\,
\cdot\,F_\beta({\mathbf{x}},\,t) \rule{0mm}{9mm}\nonumber \\
& & (i\,=\,0,\,1,\,\ldots\,{\mathcal{N}})
\rule{0mm}{8mm}
\label{realeq}
\end{eqnarray}
where $\psi\,=\,\delta s\,/\,c$. Consequently, the stationary mass and
momentum equations recovered from Eqs.~(\ref{realeq}) using the Chapman -
Enskog procedure up to second order in the Knudsen number, are
\begin{eqnarray}
\partial_\beta (\rho u_\beta) & = & \psi\partial_\beta\partial_\gamma\,
\left[\,\chi c^2\rho\delta_{\beta\gamma}\,+\,\rho u_\beta u_\gamma\,\right]
\label{masseq}\\
\partial_\beta(\rho u_\alpha u_\beta)\,+\,\partial_\alpha p_w & = &
\chi c^2\tau\partial_\beta\left[\,\rho\partial_\alpha u_\beta\,+\,
\rho\partial_\beta u_\alpha\,\right]
\rule{0mm}{8mm}  \label{momeq}\\
& + & \chi c^2\psi\left[\,
2\partial_\alpha\partial_\beta(\rho u_\beta)\,+\,\nabla^2(\rho u_\alpha)
\,\right]\, 
+ \, \rho\kappa\partial_\alpha(\nabla^2\rho) \rule{0mm}{8mm}\nonumber
\end{eqnarray}
From the mass equation (\ref{masseq}) we get the expression of the
spurious velocity which is present in the interface region
where the fluid density is not constant:
\begin{equation}
u_\beta\,=\,\frac{\,\psi\,}{\,\rho\,}\,\partial_\gamma\,\left[\,
\chi c^2\rho\delta_{\beta\gamma}\,+\,\rho\,u_\beta u_\gamma\,\right]\,
\simeq\,\frac{\,\chi c^2\psi\,}{\,\rho\,}\,\partial_\beta\rho
\label{uspurious}
\end{equation}
Thus, the spurious velocity is related to the numerical error introduced by the
first order upwind finite difference scheme. Eq.~(\ref{uspurious}) is
in good agreement with simulation results (Figure 7), especially
when the system temperature is near the critical point or when the
surface tension parameter $\kappa$ is large, i.e., when
density gradients and the magnitude of the spurious velocity in the
interface region are small.

\section*{\normalsize 4. Correction force term: reduction of the spurious
interface velocity}

To eliminate the spurious term in the mass equation (\ref{masseq}),
we consider a supplementary force term
\begin{equation}
F^{\nu}_{i\beta} \, = \,\frac{\,\psi\,}{\,\rho\,}\,\left[\,-\partial_\beta
(e_{i\gamma}A_\gamma)\,+\,(u_\gamma\partial_\beta\,+\,u_\beta\partial_\gamma)\,
A_\gamma\,\right]
\label{correctionterm}
\end{equation}
where
\begin{equation}
A_\gamma\,=\,\chi c^2\partial_\gamma\rho\,+\,\partial_\delta
(\rho u_\gamma u_\delta)
\end{equation}
After introduction of the correction force term (\ref{correctionterm}),
the LB equations (\ref{realeq}) become
\begin{eqnarray}
{\mathbf{e}}_i\,\cdot\,\nabla f_i({\mathbf{x}},\,t)\,-\,
\psi e_{i\beta}e_{i\gamma}\partial_\beta\partial_\gamma f_i({\mathbf{x}},t)
& = & - \, \frac{\,1\,}{\,\tau\,}\,\left[\, f_i({\mathbf{x}},t)
\,-\, f^{eq}_i({\mathbf{x}},t)\,\right]\,+\nonumber \\
\frac{\,1\,}{\,\chi c^2\,}\,f_i^{eq}\,
\left[\,e_{i\beta}\,-\,u_\beta({\mathbf{x}},\,t)\,\right]\,
\left[\,\,F_\beta\,+\,F^\nu_{i\beta}\,\right]
& & (i\,=\,0,\,1,\,\ldots\,{\mathcal{N}})
\rule{0mm}{7mm}
\label{realeqnew}
\end{eqnarray}

Introduction of the correction term cancels the spurious velocity term and
allows the recovery of the correct mass
equation in the stationary case. This is easy to see after calculating
the sum
\begin{eqnarray}
\sum_i\,\frac{\,1\,}{\,\chi c^2\,}\,f_i^{eq}\,(e_{i\beta}\,-\,u_\beta)\,
F_{i\beta}^{\nu} & = & \nonumber\\
\sum_i\,\frac{\,1\,}{\,\chi c^2\,}\,\frac{\,\psi\,}{\,\rho\,}\,f_i^{eq}\,
(e_{i\beta}\,-\,u_\beta)\,\left[\,-\partial_\beta
(e_{i\gamma}A_\gamma)\,+\,(u_\gamma\partial_\beta\,+\,u_\beta\partial_\gamma)\,
A_\gamma\,\right] & = & \rule{0mm}{5mm}  \label{masscancel} \\
-\,\psi\,\partial_\beta A_\gamma\delta_{\beta\gamma}\,=\,-\,
\psi\,\chi c^2\partial_\beta\,
\left[\,\partial_\beta\rho\,+\,\partial_\delta(\rho u_\beta u_\delta)\,\right]
\rule{0mm}{4mm} \nonumber
\end{eqnarray}
Moreover, the introduction of the correction force term does not alter
the momentum equation since
\begin{eqnarray}
\sum_i\,\frac{\,m\,}{\,\chi c^2\,}\,\frac{\,\psi\,}{\,\rho\,}\,f_i^{eq}\,
e_{i\alpha}(e_{i\beta}\,-\,u_\beta)\,\left[\,-\partial_\beta
(e_{i\gamma}A_\gamma)\,+\,(u_\gamma\partial_\beta\,+\,u_\beta\partial_\gamma)\,
A_\gamma\,\right] & \simeq & \nonumber\\ 
-\,\psi\,\left[\,\delta_{\alpha\beta}u_\gamma\,+\,
\delta_{\beta\gamma}u_\alpha\,+\,\delta_{\alpha\gamma}U_\beta\,\right]\,
\partial_\beta A_\gamma\,+\,\psi\delta_{\alpha\beta}\,\left[\,
u_\gamma\partial_\beta\,+\,u_\beta\partial_\gamma\,\right]\,A_\gamma & + &
\rule{0mm}{4mm}\label{momcancel}\\
\psi\delta_{\alpha\gamma}\partial_\beta A_\gamma  & = & 0 \nonumber
\rule{0mm}{6mm} \nonumber
\end{eqnarray} 

In the $1D$ case there is no need for Cartesian indices and
the expression (\ref{correctionterm}) of the correction
force term becomes
\begin{equation}
F_{i}^{\nu}\,=\,-\,\frac{\,\psi\,}{\,\rho\,}\,(\,e_i\,-\,2u\,)\,\left[\,
\chi c^2\nabla^2\rho\,+\,\nabla^2(\rho u^2)\,\right]
\label{f1d}
\end{equation}
The following formula is used to compute the effect of the Laplace operator
on a function $f$ (e.g., $\rho$ or $\rho u^2$)
defined in the nodes $x$ of the $1D$ lattice:
\begin{equation}
\nabla^2 f({\mathbf{x}},t)\,=\,\frac{\,2\,}{\,\chi(\delta s)^2\,}\,\left[\,
\sum_{i=0}^{i={\mathcal{N}}}\,
w_i\,f({\mathbf{x}}+{\mathbf{e}}_i\delta s/c,t)\,-\,f({\mathbf{x}},t)
\,\right]
\label{lapformula}
\end{equation}

\begin{figure}
\centerline{\includegraphics[width=100mm]{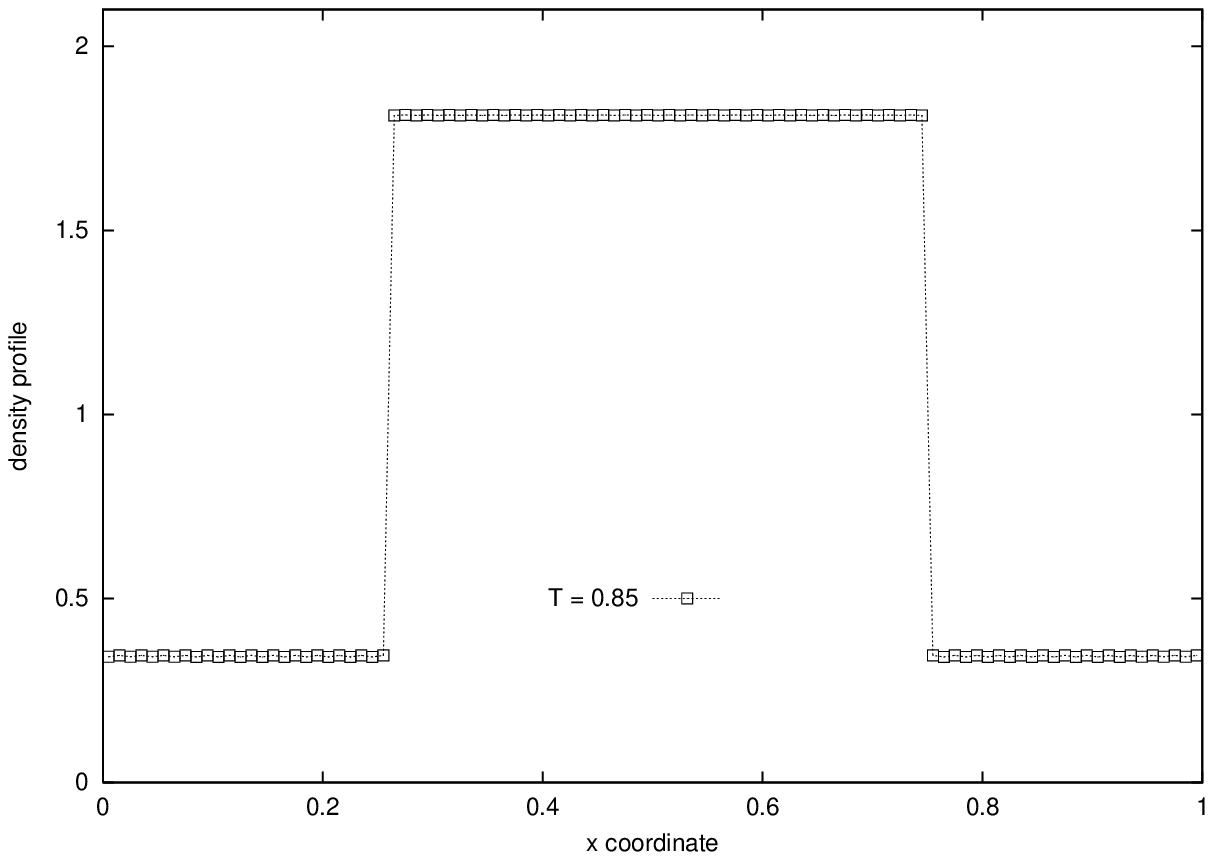}}

\centerline{a: density profile}

\label{correctedprofiles}
\caption{Profiles recovered using the correction force
term (\protect{\ref{correctionterm}}) for $T\,=\,0.85$ and $\kappa\,=\,0$.}
\end{figure}

\addtocounter{figure}{-1}
\begin{figure}
\centerline{\includegraphics[width=100mm]{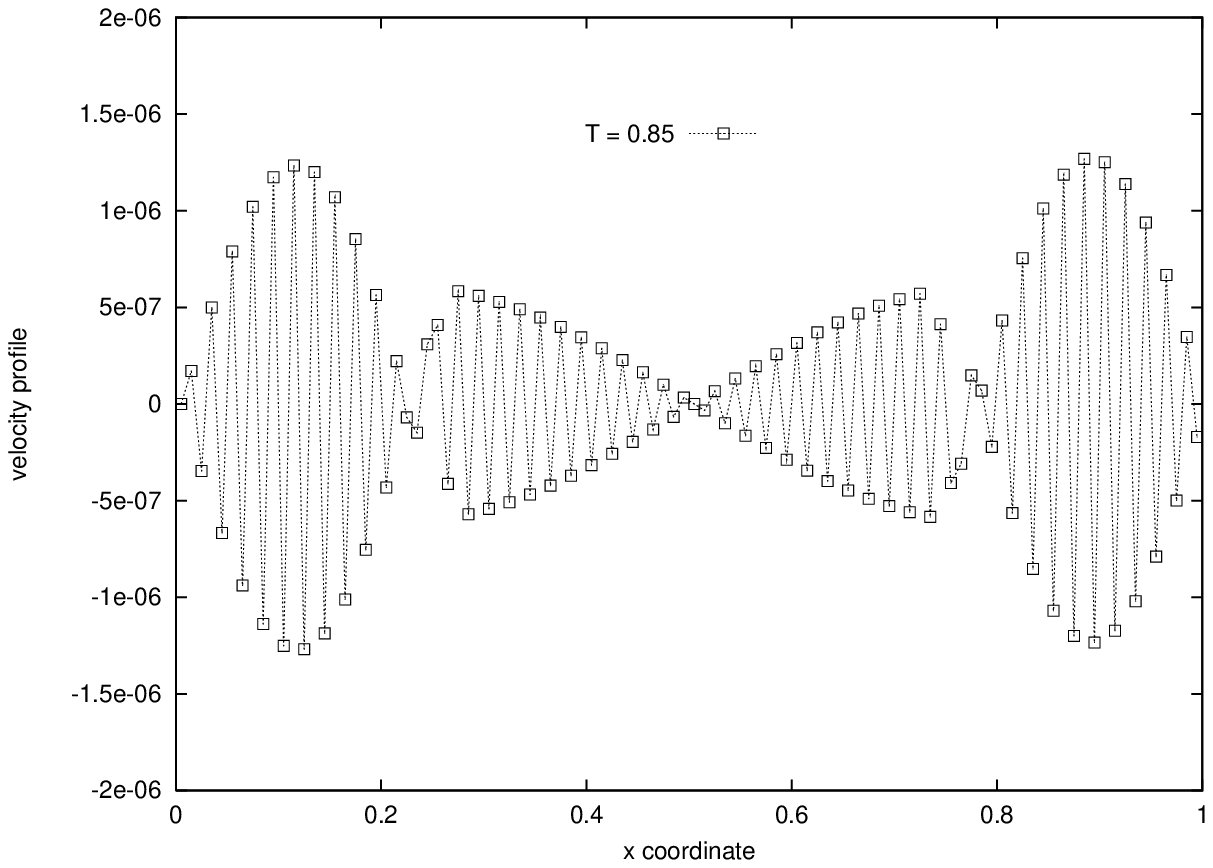}}

\centerline{b: velocity profile}

\centerline{\includegraphics[width=100mm]{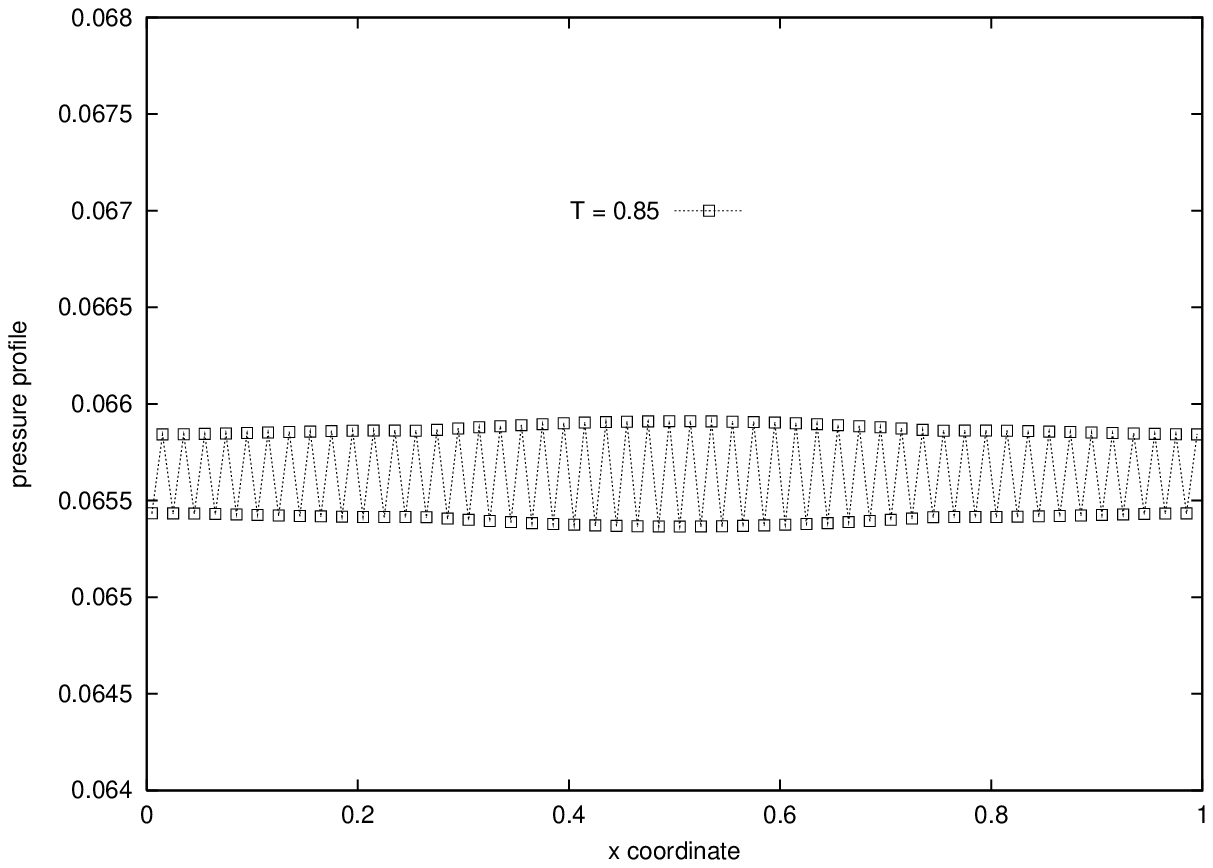}}

\centerline{c: pressure profile}

\label{correctedprofilescont}
\caption{{\emph{(cont'd)}} Profiles recovered using the correction force
term (\protect{\ref{correctionterm}})
for $T\,=\,0.85$ and $\kappa\,=\,0$.}
\end{figure}

\begin{figure}
\centerline{\includegraphics[width=100mm]{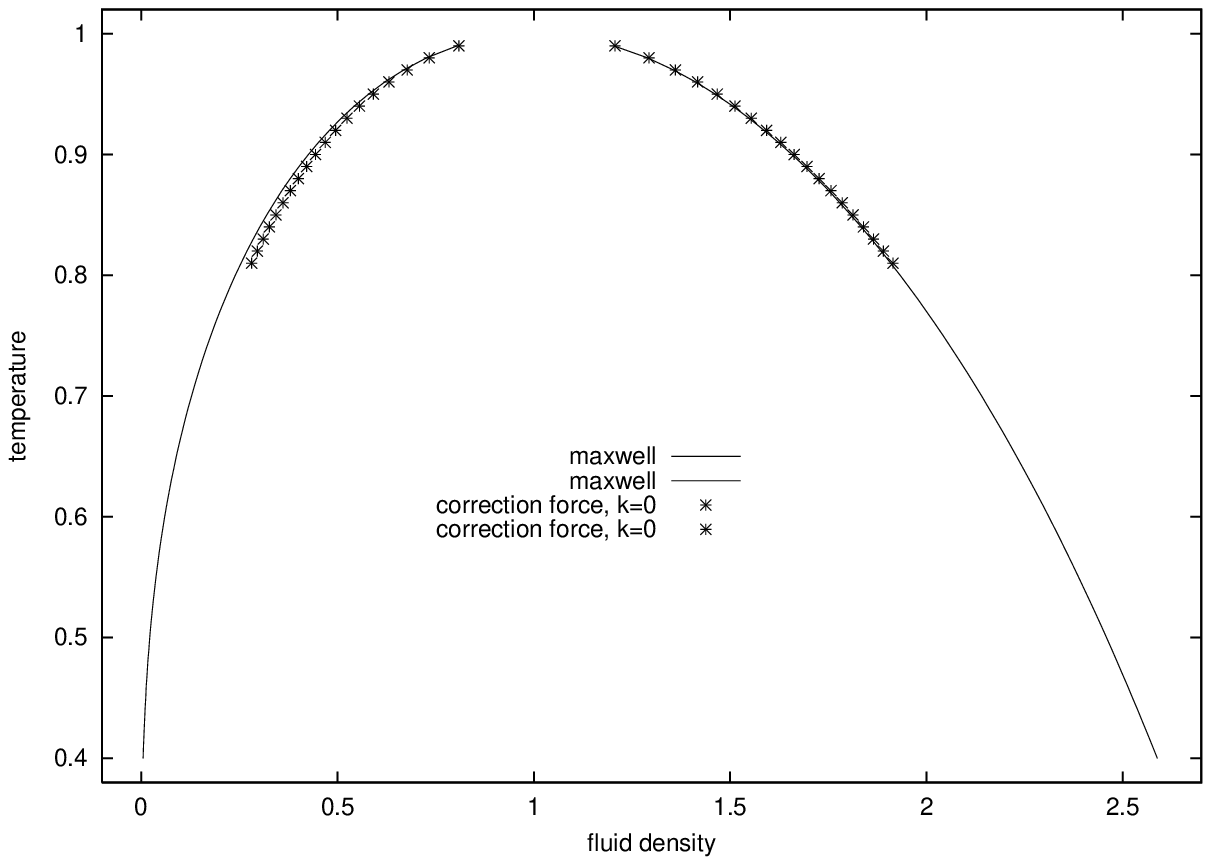}}

\centerline{a: $\kappa\,=\,0$}

\centerline{\includegraphics[width=100mm]{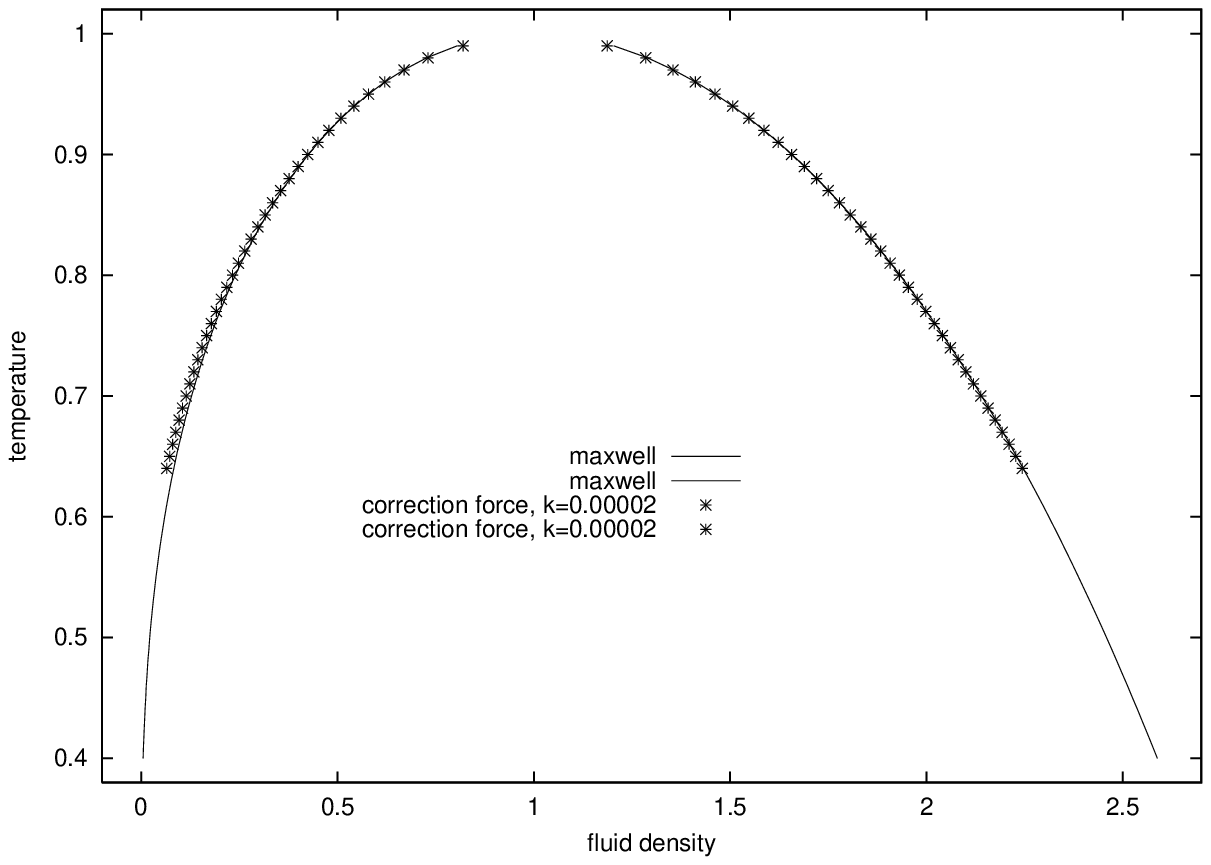}}

\centerline{b: $\kappa\,=\,0.00002$}

\label{phasediagcorrected}
\caption{Phase diagram recovered with the correction force
term (\protect{\ref{correctionterm}}).}
\end{figure}

\addtocounter{figure}{-1}
\begin{figure}
\centerline{\includegraphics[width=100mm]{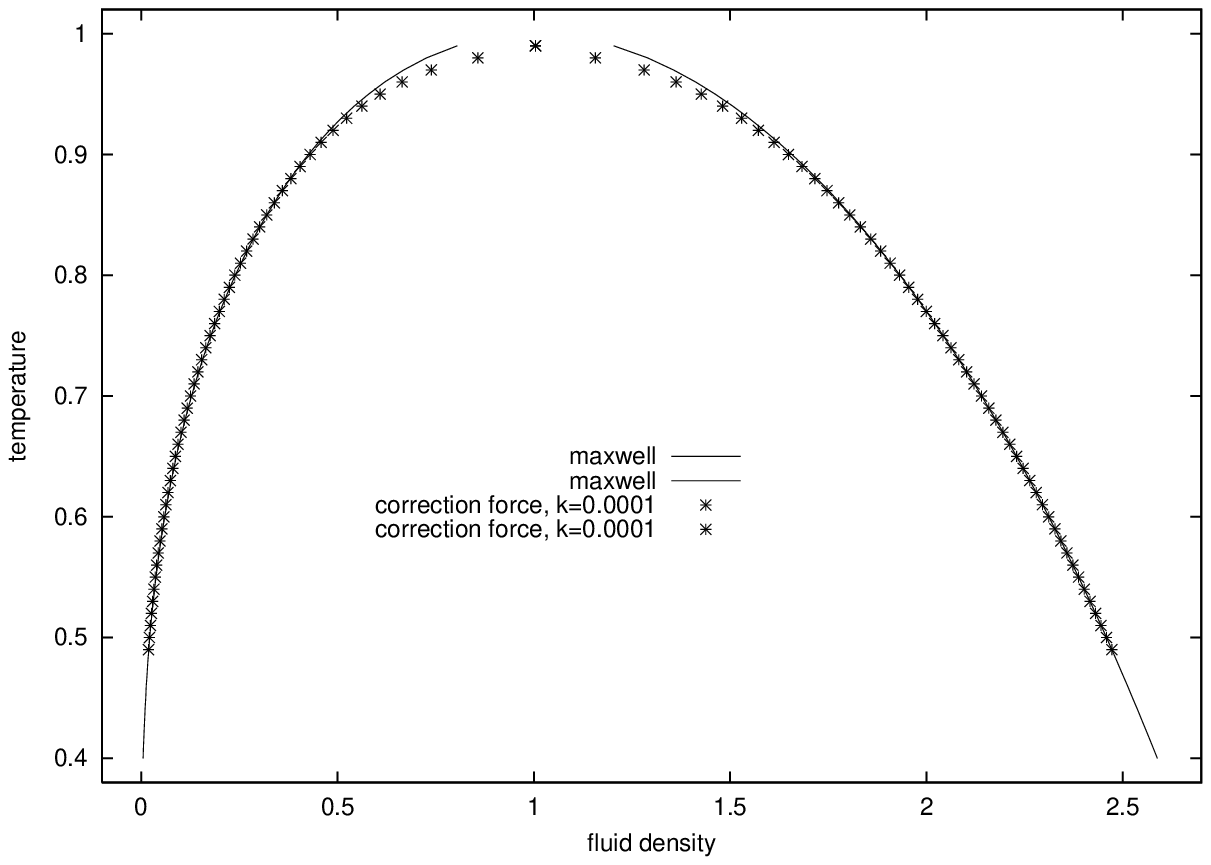}}

\centerline{c: $\kappa\,=\,0.0001$}

\label{phasediagcorrectedcont}
\caption{{\emph{(cont'd)}} Phase diagram recovered with the correction force
scheme (\protect{\ref{correctionterm}}).}
\end{figure}

Simulations done with the correction force term (\ref{correctionterm})
included in the LB evolution
equations give sharp density profiles for $\kappa\,=\,0$
(Figure 8a), as expected. The spurious velocity is no longer present
in the interface region. A wiggly profile of the local
velocity is still observed (Figure 8b), but the amplitude of the
oscillations are several orders of magnitude smaller than the
typical value of the spurious velocity arising in the interface
region when using the uncorrected upwind scheme. Also the pressure profile
recovered for
$\kappa\,=\,0$ is quite flat on the whole lattice, although wiggles
of very small amplitude are still present (Figure 8c).

As previously
observed during simulations with the uncorrected upwind scheme,
increasing the surface tension always helps to stabilize the system for
lower values of the temperature when the correction force term
is considered in the LB evolution equations (Figure 9).
However, the phase diagrams recovered
using the corrected force term (Figure 9) are closer to theoretical results
than the phase diagrams derived with the bare upwind scheme (Figure 6).
For large values of the surface tension parameter $\kappa$, the overlapping
of the left and right interfaces is still present
for temperatures close to the critical point, but the plateau
is clearly defined in both high and low density phases for lower temperatures
(Figure 10). The velocity profile recovered
for $T\,=\,0.50$ and $\kappa\,=\,0.0001$ is
similar to the one shown in Figure 6b, but the amplitude of the oscillations
is significantly reduced ($2.0\times 10^{-14}$) because of the stabilization
effect of the surface tension.

\newpage

\begin{figure}
\centerline{\includegraphics[width=100mm]{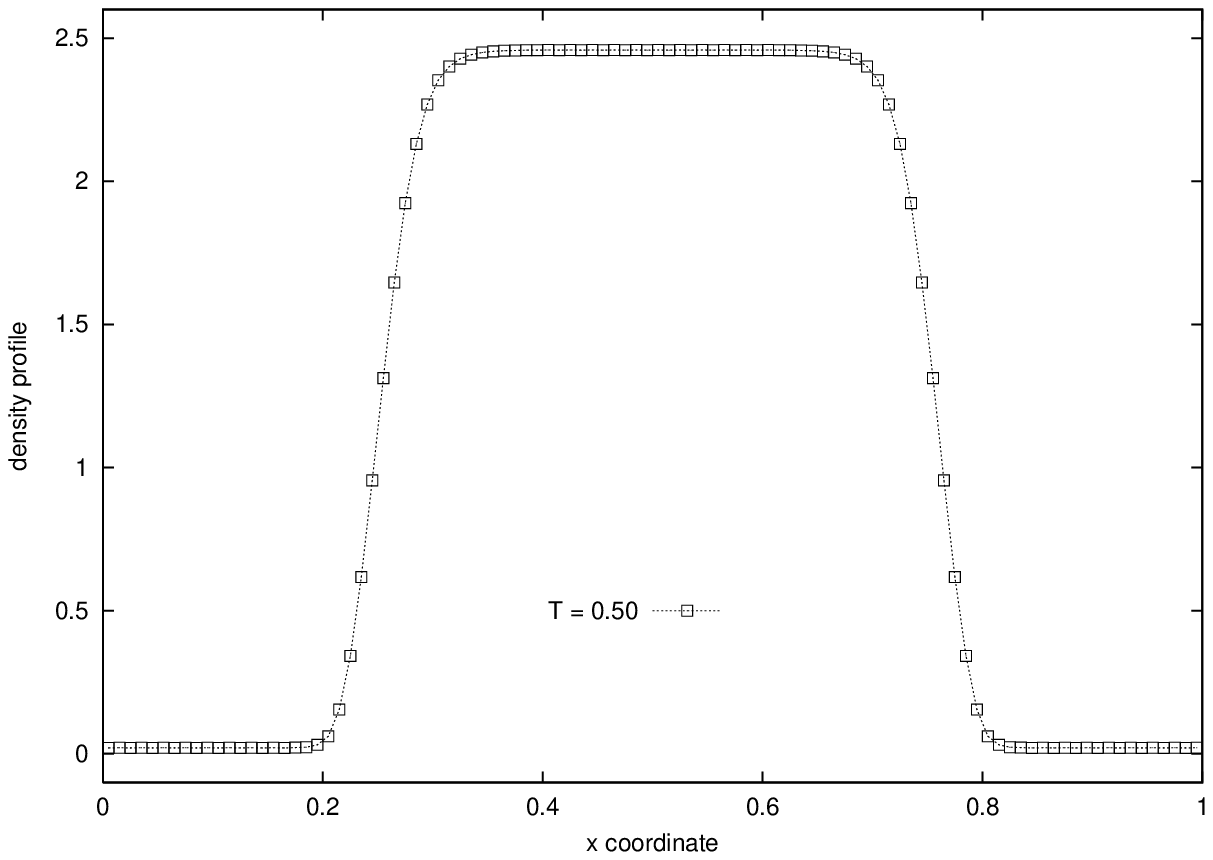}}

\centerline{a: density profile}

\centerline{\includegraphics[width=100mm]{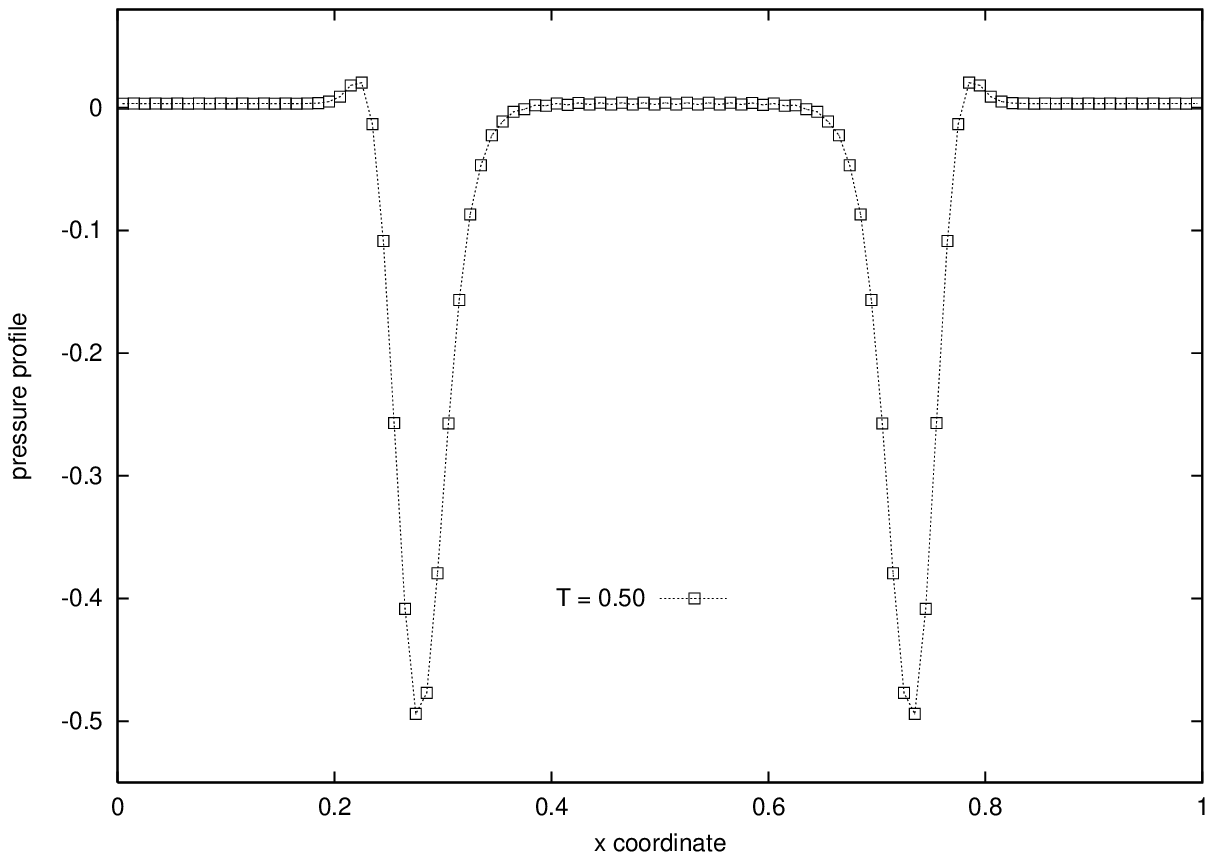}}

\centerline{b: pressure profile}

\label{correctedprofilesT50}
\caption{{\emph{(cont'd)}} Profiles recovered using the correction force
term (\protect{\ref{correctionterm}}) for $T\,=\,0.50$ and
$\kappa\,=\,0.0001$.}
\end{figure}

\section*{\normalsize 5. Conclusions}

The first order
upwind finite difference scheme used for the operator ${\mathbf{e}}_i
\cdot\nabla$ in the LB evolution equations (\ref{fdlbeqadimfin})
generates the spurious velocity in the interface region of
liquid - vapor systems. The magnitude of the spurious velocity is
related to the density gradient through Eq.~(\ref{uspurious}). Consequently,
the spurious velocity is reduced by the surface tension.

A correction force term (\ref{correctionterm})
is suggested to reduce the spurious velocity. Introduction of this term
in the LB evolution equations allows to get sharp interfaces when
the value of the surface tension parameter vanishes. Moreover,
the phase diagram of the liquid - vapor system
becomes closer to the theoretical one derived
using the Maxwell construction.


\end{document}